\documentclass[aps,twocolumn,groupedaddress,superscriptaddress,showpacs]{revtex4}
\usepackage{graphicx}
\usepackage{colortbl}

\makeatletter
\def\btt#1{\texttt{\@backslashchar#1}}%
\DeclareRobustCommand\bblash{\btt{\@backslashchar}}%
\makeatother

\begin{document}

%\baselineskip=2\normalbaselineskip
%↑１行空きにする場合

\title{Mottness and spin liquidity in a doped organic superconductor $\kappa$-(BEDT-TTF)$_4$Hg$_{2.89}$Br$_8$} 

\author{H. Oike}
\email{oike@ap.t.u-tokyo.ac.jp}
{\affiliation{PRESTO, Japan Science and Technology Agency (JST), Kawaguchi 332-0012, Japan}
\affiliation{Department of Applied Physics, The University of Tokyo, Tokyo 113-8656, Japan}

\author{H. Taniguchi}
{\affiliation{Graduate School of Science and Engineering, Saitama University, Saitama 338-8570, Japan}

\author{K. Miyagawa}
{\affiliation{Department of Applied Physics, The University of Tokyo, Tokyo 113-8656, Japan}

\author{K. Kanoda}
\email{kanoda@ap.t.u-tokyo.ac.jp}
\affiliation{Department of Applied Physics, The University of Tokyo, Tokyo 113-8656, Japan}
\affiliation{Max Planck Institute for Solid State Research, Heisenbergstrasse 1, 70569 Stuttgart, Germany}
\affiliation{Physics Institute, University of Stuttgart, Pfaffenwaldring 57, D-70569 Stuttgart, Germany}
\affiliation{Department of Advanced Materials Science, University of Tokyo, Kashiwanoha 5-1-5, Kashiwa 277-8561 Chiba, Japan}

\date{\today}
\begin{abstract}
It has been more than 40 years since superconductivity was discovered in organic conductors, and the way scientists view organic superconductors has changed over time. At first, the fact that organic conductors exhibit superconductivity was a novelty in itself, and subsequently it was shown that behind the superconductivity is the physics of electron correlation, which has been a focus in condensed matter physics at large. Amid the remarkable development of correlation physics, the unique characteristics of organic conductors, e.g., a variety of lattice geometries and the highly compressible feature, led to the elucidation of fundamental principles and the finding of new phenomena, such as bandwidth-controlled Mott transitions and possible quantum spin liquids. However, most organic superconductors have commensurate band fillings, such as a half or a quarter, whereas inorganic superconductors, such as high-$T_{\rm c}$ cuprates and iron-based superconductors, have often been investigated under the variation of their band fillings. Thus, the physical linkage between organic and inorganic superconductors has remained unresolved. In this review article, we focus on the layered nonstoichiometric superconductor, $\kappa$-(BEDT-TTF)$_4$Hg$_{2.89}$Br$_8$, which is exceptional among organic conductors in that the nonstoichiometry serves as doping to a half-filled band. Moreover, the strong correlation of electrons and a geometrically frustrated triangular lattice make this system exhibit the unique phenomena involved in Mottness, spin liquidity, and superconductivity, which are key concepts of correlated electron physics. This review will summarize what we learned from the pressure study of $\kappa$-(BEDT-TTF)$_4$Hg$_{2.89}$Br$_8$ and how they relate to the extensively studied issues in inorganic materials.
\end{abstract}

\pacs{xxxx}

\maketitle
\tableofcontents

%***********************************************************************
\section{Introduction}
The 1980s was a period of discovery of various superconducting organic crystals, such as TMTSF-based compounds and BEDT-TTF-based compounds, where TMTSF and BEDT-TTF represent tetramethyltetraselenafulvalence and bis(ethylenedithio)tetrathiafulvalene, respectively \cite{bechgaard1980properties,jerome1980superconductivity,jerome1982organic,parkin1983superconductivity,yagubskii1985coexistence,kobayashi1986crystalicl2,kobayashi1986new,kobayashi1986crystal,kobayashi1987crystal,kajita1987new,lyubovskaya1987organic,lyubovskaya1987superconductivity,urayama1988new,williams1991organic}. Among them is the paper on superconductivity in $\kappa$-(BEDT-TTF)$_4$Hg$_{2.89}$Br$_8$ ($\kappa$-Hg$_{2.89}$Br$_8$), published in 1987 \cite{lyubovskaya1987organic,lyubovskaya1987superconductivity}, a year after the discovery of high-temperature superconductivity in cuprates \cite{bednorz1986possible}. From a modern perspective, both superconductors emerge when doping a Mott insulator \cite{lee2006doping,taniguchi2007anomalous,oike2015pressure}, and thus, it is reasonable that we imagine a common physics behind them. However, these discoveries were probably seen as events in different fields at that time. The paper on high-$T_{\rm c}$ cuprates immediately attracted tremendous attention and sparked the study of strongly correlated electron systems. On the other hand, it took some time before the background for studying $\kappa$-Hg$_{2.89}$Br$_8$ could be formed. 

In the 1990s, it began to be recognized that organic conductors are suitable for studying strongly correlated electron systems \cite{kino1995electronic,kanoda1997electron,kanoda1997hyperfine}. Although the crystal structures of molecular materials are complex, the conduction bands have a simple structure that is well reproduced by overlapping frontier orbitals of molecules as reproduced with extended H\"{u}ckel and tight-binding approximations\cite{kino1995electronic}. In addition to the simplicity of the band structure, the pressure tunability of the lattice constants is favorable for investigating parameter-dependent physical properties. Pressure changes the orbital overlap together with the lattice parameters, and consequently affects the key parameters of correlated electron physics such as the kinetic energy of the conduction electrons and the interaction strength between electrons. Thus, pressure studies on $\kappa$-type BEDT-TTF compounds have revealed a concrete picture of a bandwidth-controlled Mott transition in half-filled systems, namely, non-doped systems. 

The development of Mott physics in $\kappa$-type BEDT-TTF compounds has provided the basis for $\kappa$-Hg$_{2.89}$Br$_8$ to be viewed as a doped Mott insulator. However, this did not immediately lead to the development of the physics of doped systems in organic conductors because of the difficulties of experiments. $\kappa$-Hg$_{2.89}$Br$_8$ often exhibits crystalline phases other than the targeted phase \cite{li1998room}, and it has been difficult to find a method to synthesize the targeted crystalline phase with good reproducibility. Indeed, only a few groups, including one of the authors, were successful in synthesizing $\kappa$-Hg$_{2.89}$Br$_8$ after the discovery by Lyubovskaya et al \cite{lyubovskaya1987organic,li1998room,taniguchi2007anomalous}. Moreover, microcracks often appear in the material upon cooling, making it difficult to quantify electrical conduction phenomena at low temperatures. Perhaps it was also difficult to find an analytical method to reveal new physics from the experimental results of $\kappa$-Hg$_{2.89}$Br$_8$.

In the 2000s, experimental indications of a quantum spin liquid (QSL) in $\kappa$-(BEDT-TTF)$_2$Cu$_2$(CN)$_3$ led to further interest in $\kappa$-type BEDT-TTF compounds \cite{shimizu2003spin}. The $\kappa$-type compounds have isosceles triangular lattices and exhibit antiferromagnetic (AF) orders in the Mott insulating states \cite{miyagawa1995antiferromagnetic}; however, $\kappa$-(BEDT-TTF)$_2$Cu$_2$(CN)$_3$, a Mott insulator with a nearly right triangular lattice, was found to have no magnetic order and was suggested to have gapless excitations reminiscent of Fermi degeneracy at low temperatures although its ground state is still under debate \cite{shimizu2003spin,shimizu2006emergence,yamashita2008thermodynamic,yamashita2009thermal}. Amid the longstanding quest for QSLs on geometrically frustrated lattices, this experimental indication triggered the discoveries of other QSL candidates \cite{itou2008quantum,isono2014gapless,shimizu2016pressure} and consequently made $\kappa$-Hg$_{2.89}$Br$_8$ unique because it also has a similar triangular lattice and thus may host a doped QSL. Doped QSL was theoretically proposed as the origin of high-$T_{\rm c}$ superconductors \cite{anderson1987resonating}, but their parent materials are antiferromagnets. Thus, $\kappa$-Hg$_{2.89}$Br$_8$ should play a key role in understanding the superconductivity emerging from a QSL.

In this review, we describe how the study of $\kappa$-Hg$_{2.89}$Br$_8$ has opened up new aspects of strongly correlated electron systems. Analogous to the bandwidth-controlled Mott transition in a half-filled system, how do electron correlations give rise to nonperturbative effects in the doped system? Although charge is not mobile in QSL states of a half-filled system, how does carrier doping superimpose charge degrees of freedom over the exotic spin states? How does superconductivity emerge in these situations? Although there is still much to be learned about these diverse phenomena from $\kappa$-Hg$_{2.89}$Br$_8$, this review summarizes what we have found in the doped Mott insulator.

In Chapter 2, we provide an overview of the Mott transition, QSL and superconductivity exhibited by $\kappa$-type BEDT-TTF compounds for a clearer understanding of $\kappa$-Hg$_{2.89}$Br$_8$. Because many review articles are available on these topics in $\kappa$-type BEDT-TTF compounds \cite{kanoda1997electron,miyagawa2004nmr,kanoda2006metal,kanoda2011mott, zhou2017quantum,pustogow2022thirty,wang2023organic}, we provide a brief overview of the fundamentals and present status relevant to $\kappa$-Hg$_{2.89}$Br$_8$. Chapter 3 describes the synthesis and crystal structure of $\kappa$-Hg$_{2.89}$Br$_8$, emphasizing its uniqueness. Chapters 4-6 describe the unusual electronic properties of $\kappa$-Hg$_{2.89}$Br$_8$ in detail and discuss doping effects on the Mott transition, QSL, and superconductivity, respectively. Chapter 7 summarizes this review.

\section{Electronic states of $\kappa$-type BEDT-TTF compounds}
A characteristic feature of $\kappa$-type BEDT-TTF compounds is that materials with similar crystal structures exhibit a wide variety of electronic states. In all compounds, electrons in the HOMO (highest occupied molecular orbital) of BEDT-TTF molecules are solely responsible for the electrical conduction and magnetism. Therefore, although the phenomena are diverse, they can be attributed to the behavior of electrons in the same type of orbital. This feature has worked effectively in developing the physics of strongly correlated electron systems. This chapter describes the basic properties of non-doped $\kappa$-type BEDT-TTF compounds.

\subsection{Crystal and band structures}
Fig. 1 shows the crystal structure of $\kappa$-(BEDT-TTF)$_2$Cu$_2$(CN)$_3$ as an example of $\kappa$-type BEDT-TTF compounds, which have a layered structure composed of anion layers and BEDT-TTF layers (Fig. 1(a)) \cite{geiser1991superconductivity}. The anion layers pull out one electron from every two BEDT-TTF molecules to form a closed shell, and consequently, the HOMOs of BEDT-TTF molecules become an open shell. Therefore, anion layers are nonmagnetic and insulating, similar to a block layer of high-$T_{\rm c}$ cuprates, and BEDT-TTF layers play a main role in correlated electron physics. Fig. 1(b) shows a $\kappa$-type molecular arrangement in a BEDT-TTF layer. Because the hybridization between atomic orbitals in a molecule are much larger than inter-molecular transfer integrals \cite{mori1999structural}, the molecular orbitals can be regarded as minimum units of the electronic structure. This hierarchical structure gives a rather simple view of $\kappa$-type BEDT-TTF compounds although the atomic arrangement in real space is complicated. 

%%%%%%%%%%%%%%%%%%%Fig1
\begin{figure}
\includegraphics[width=86mm]{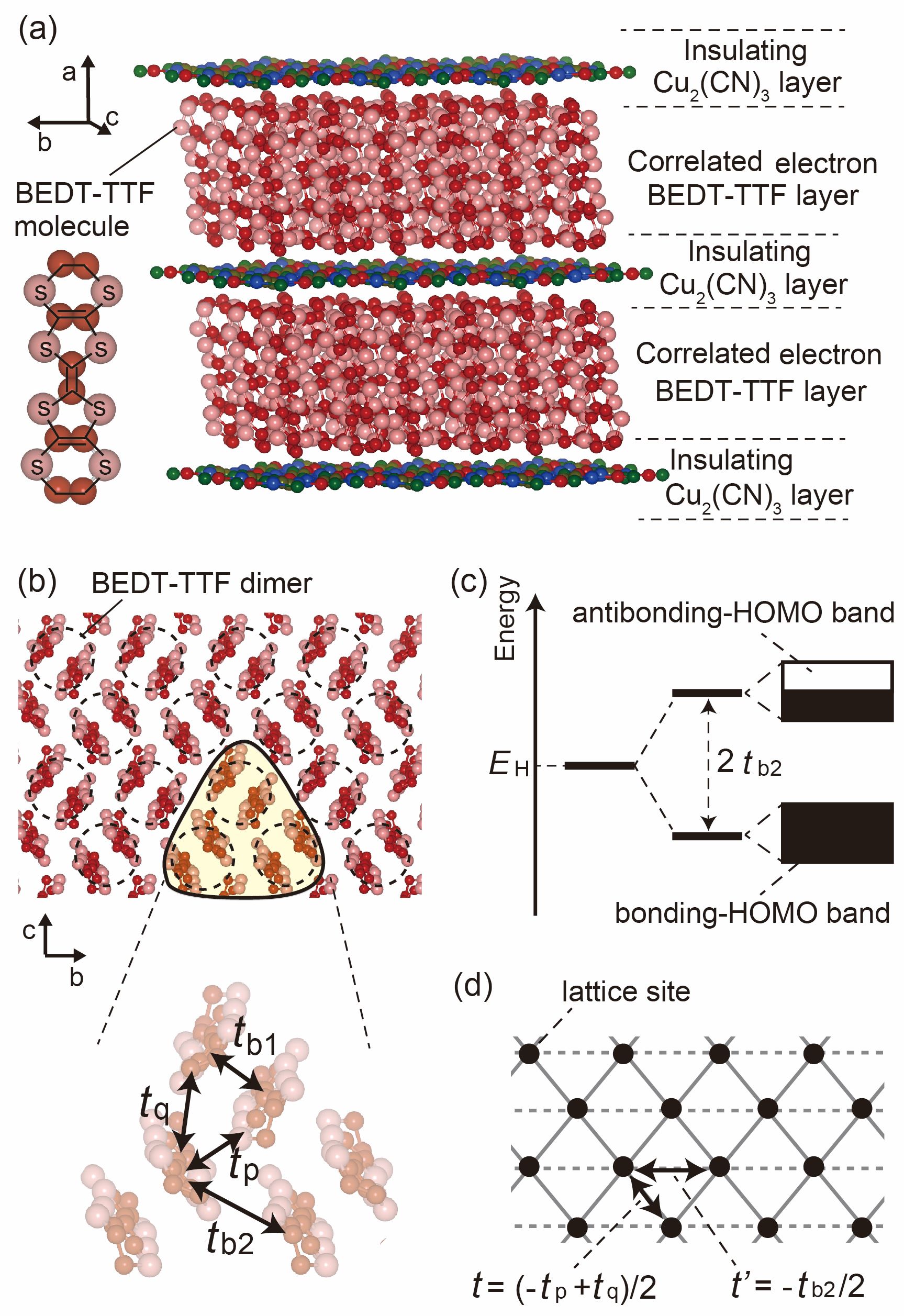}
\caption{\label{Fig1}
(a) Crystal structure of $\kappa$-(BEDT-TTF)$_2$Cu$_2$(CN)$_3$. (b) $\kappa$-type molecular arrangement in a BEDT-TTF layer. According to the extended H\"{u}ckel approximations, the conduction and valence bands are mainly represented by four inter-molecular transfer integrals, $t_{\rm b1}$, $t_{\rm b2}$, $t_{\rm p}$, and $t_{\rm q}$  \cite{kino1995electronic}. (c) Schematic representation of energy splitting of HOMO to form a conduction band. (d) Isosceles trianglar lattice formed by BEDT-TTF dimers. Inter-dimer transfer integrals $t$ and $t'$ are represented by $t_{\rm b2}$, $t_{\rm p}$, and $t_{\rm q}$.}
\end{figure}
%%%%%%%%%%%%%%%%%%%Fig1

Based on the extended H\"{u}ckel and tight binding approximations, the dimer formation of two molecules makes the HOMOs split into bonding and antibonding orbitals. When the intra-dimer transfer integral $t_{\rm b1}$ is several times larger than the inter-dimer transfer integrals, the conduction band composed of antibonding HOMOs is half-filled (Fig. 1(c)) \cite{kino1995electronic,hotta2003classification}. Although weak dimerization results in a quarter-filled conduction band \cite{drichko2014metallic, hassan2018evidence, hassan2020melting}, which is composed of both bonding and antibonding HOMOs, this review focuses on compounds with relatively strong dimerization. Then, antibonding HOMOs form a triangular lattice characterized by inter-dimer transfer integrals $t$ and $t'$ (Fig.1(d)).

%%%%%%%%%%%%%%%%%%%Fig2
\begin{figure}
\includegraphics[width=86mm]{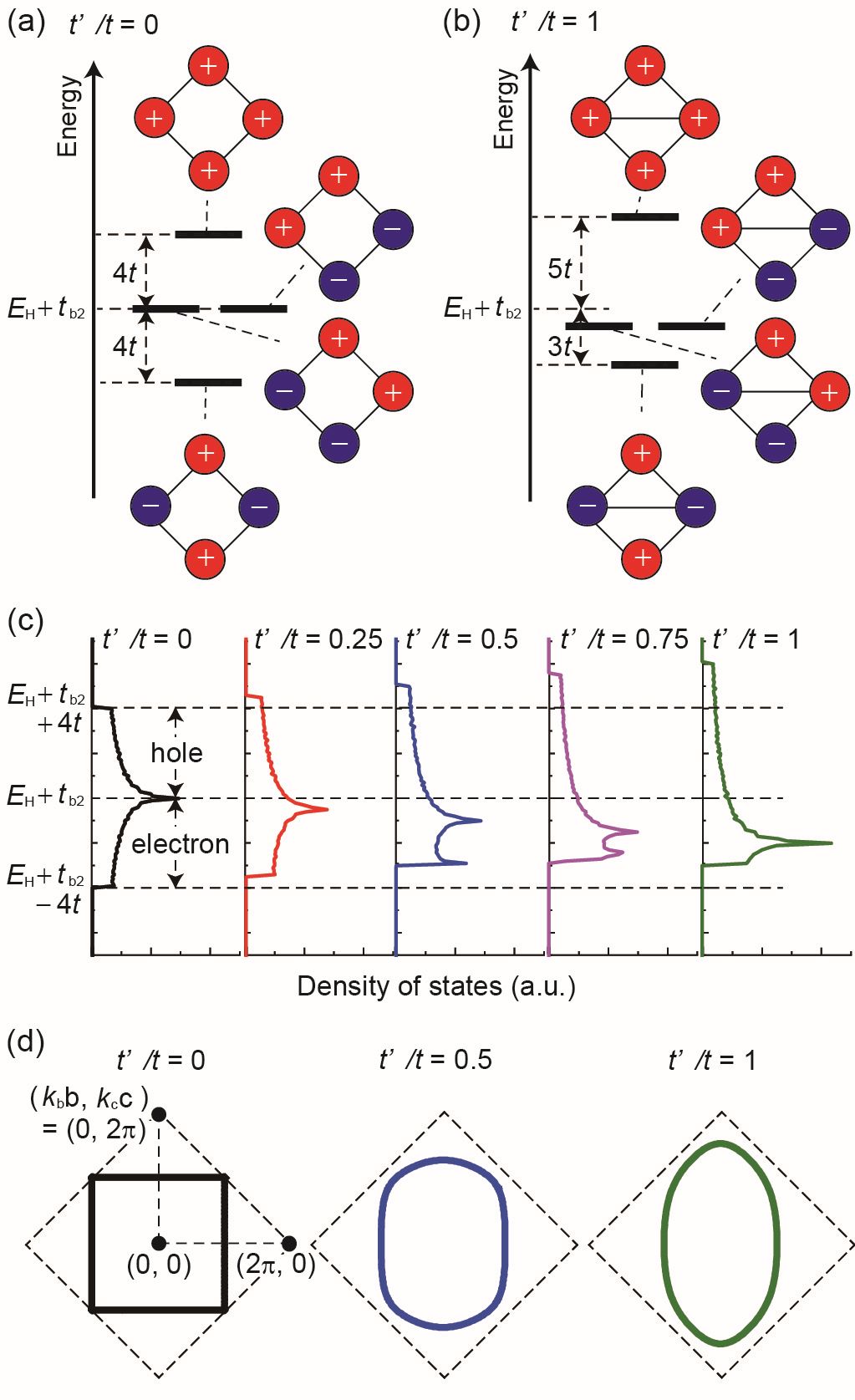}% Here is how to import EPS art
\caption{\label{fig}
(a) Energy levels of four dimers with four bonds of a transfer integral $t$. A circle indicates a dimer. Because of the anti-bonding nature of dimers, the inter-dimer bond is stabilized when the signs of the phase of dimer orbitals are opposite to each other, such as sigma bonds formed from p orbitals. (b) Energy levels of four dimers with an additional diagonal bond of a transfer integral $t'$. (c) Density of states of the isosceles triangular lattice formed by inter-dimer transfer integrals $t$ and $t'$. (d) Fermi surfaces of the isosceles triangular lattice at half filling. $b$ ($c$) and $k_{\rm_b}$ ($k_{\rm_c}$) represent the lattice constant and wavenumber along the b axis (c axis), respectively. When $t'/t = 0$, the Fermi level is at a van Hove singularity. The finite value of $t'/t$ decreases the energy level of the van Hove singularity, and consequently, the Fermi surface becomes hole-like.}
\end{figure}
%%%%%%%%%%%%%%%%%%%Fig2

Because of the anti-bonding nature of the dimer orbital, an inter-dimer bonding is stabilized when their wave functions have opposite signs (Fig.2(a)). When $t'$ is zero, the lattice is identical to a square lattice, whose Fermi surface is also square at half-filling with particle-hole symmetry (Fig.2(c,d)). On a triangular lattice, however, there are no configurations in which all inter-dimer bonds are stabilized (Fig.2(b)). The geometrical frustration of the phase factor leads to the degenerative nature of the bottom half of the conduction band for finite $t'$ values and breaks particle-hole symmetry (Fig.2(c)), resulting in a hole-like Fermi surface at half-filling (Fig.2(d)). This tight-binding picture can approximate the first-principal band structure calculation \cite{ching1997electron}. The asymmetry may be related to the difference between hole-doped and electron-doped Mott insulators \cite{kawasugi2016electron, watanabe2019mechanism, kawasugi2019non, kawasugi2021simultaneous}.

\subsection{Bandwidth-controlled Mott transition}
In the above tight-binding picture, we regard a dimer as a site, and thus, the onsite Coulomb repulsion $U$ means an energy increase due to the occupation of two electrons (or two holes in the present case) on a dimer. Therefore, the value of $U$ is different from an on-molecular repulsion $U_{\rm mol}$, which is approximately 4 eV without screening \cite{ducasse1997valence}. Because $U_{\rm mol}$ is much larger than $t_{\rm b1}$, $U$ is approximately given by 2$t_{\rm b1}$ as follows. Fig. 3 depicts two configurations: one is two “singly occupied” dimers, both of which have one hole, or equivalently three electrons (Fig.3(a)), and the other is a pair of an “empty” dimer with no hole (four electrons) and a “doubly occupied” dimer with two holes (two electrons) (Fig.3(b)). $U$ is defined as the energy difference between these two configurations. In the former configuration, the energy equals 6$E_{\rm H}$ – 2$t_{\rm b1}$ where $E_{\rm H}$ is the energy level of the HOMO. In the latter configuration, the energy is equal to 6$E_{\rm H}$ when the second order of $t_{\rm b1}/U_{\rm mol}$ is neglected. Then, the energy difference between two states nearly equals 2$t_{\rm b1}$, meaning that $U$ is approximated to be 2$t_{\rm b1}$. Since 2$t_{\rm b1}$ is comparable to the bandwidth $W$, $\kappa$-type BEDT-TTF compounds can be viewed as strongly correlated electron systems.

%%%%%%%%%%%%%%%%%%%Fig3_wide
\begin{figure*}
\includegraphics[width=172mm]{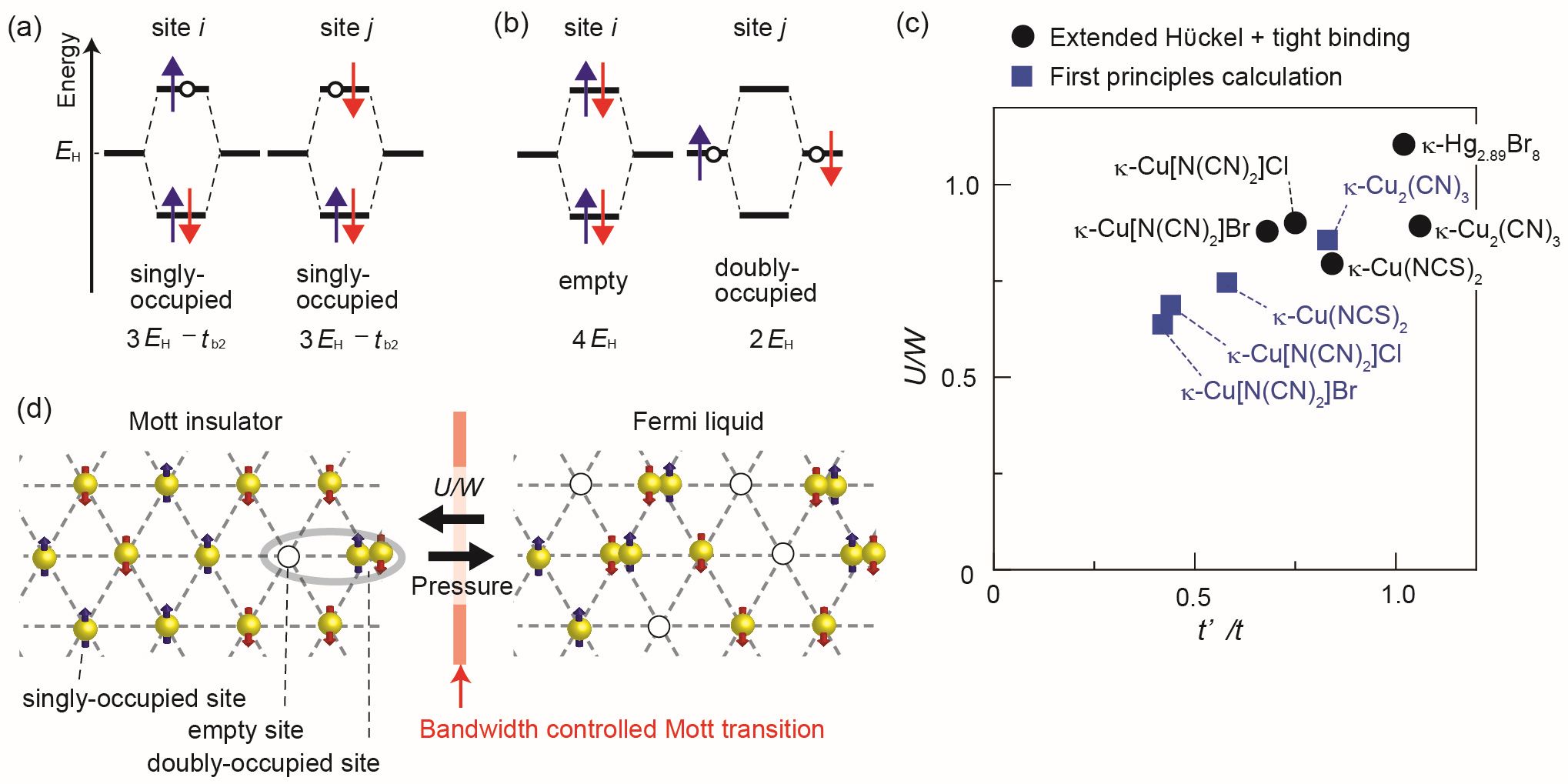}% Here is how to import EPS art
\caption{\label{fig:wide}
(a) Energy levels of two ``singly-occupied" dimers. The arrows and circles represent electrons and holes, respectively. Because the site occupancy is defined as the number of holes at a dimer in the present case, a singly occupied site has three electrons in the HOMOs. (b) A pair of an ``empty" dimer and a ``doubly occupied" dimer. If two electrons occupy a bonding orbital in a doubly-occupied site, the energy equals 2$E_{\rm H}$ + $U_{\rm mol}$ – 2$t_{\rm b1}$, which is much larger than 2$E_{\rm H}$ because of the strong on-molecular repulsion betweem holes $U_{\rm mol}$. Closed-shell molecules do not normally form hybridized orbitals, but immediately after changing from a singly-occupied site to an empty site by charge transfer with other sites, the electrons are considered to occupy the hybridized orbitals. (c) Schematic figure of the bandwidth-controlled Mott transition. Because double occupancy is strongly prohibited in a Mott insulating state, ``empty"  and ``doubly occupied" sites are created as a bounded pair. In a Fermi liquid state, ``empty"  and ``doubly occupied" sites can be unbound, resulting in metallic conduction. (d) The values of $U/W$ and $t'/t$ for $\kappa$-type BEDT-TTF compounds obtained by the extended H\"{u}ckel and tight binding approximation \cite{mori1984intermolecular,mori1999structural} and first-principle calculations \cite{kandpal2009revision}. }
\end{figure*}
%%%%%%%%%%%%%%%%%%%Fig3_wide

A Mott insulating state emerges because of electron correlation in a half-filled system, where the number of electrons in the conduction band equals that of the lattice sites. When $U$ exceeds the energy gain of forming a conduction band, the double occupancy is strongly prohibited, and then every electron becomes localized at each lattice site, resulting in the insulating state. Such Mott insulating states appear in a wide variety of materials, such as oxides, chalcogenides, organic crystals, and more recently twisted bilayer graphene \cite{imada1998metal,kanoda1997electron,kanoda2011mott, kim2018spectroscopic, andrei2020graphene}. Because a large $U/W$ is necessary for Mott insulating states to emerge, a reduction in $U/W$ results in a phase transition to a metallic state, which is referred to as a bandwidth-controlled Mott transition (Fig. 3(c)).

Fig. 3(d) shows the values of $U/W$ and $t'/t$ for $\kappa$-type BEDT-TTF compounds with a half-filled conduction band, $\kappa$-(BEDT-TTF)$_2$Cu(NCS)$_2$, $\kappa$-(BEDT-TTF)$_2$Cu[N(CN)$_2$]Br, $\kappa$-(BEDT-TTF)$_2$Cu[N(CN)$_2$]Cl, and $\kappa$-(BEDT-TTF)$_2$Cu$_2$(CN)$_3$, together with those for the doped compound $\kappa$-Hg$_{2.89}$Br$_8$ \cite{mori1984intermolecular,mori1999structural,kandpal2009revision}. The values of $U/W$ and $t'/t$ are obtained by two methods: one is the extended H\"{u}ckel and tight binding approximation \cite{mori1984intermolecular,mori1999structural}, and the other is based on first-principle band structure calculations \cite{kandpal2009revision}. Although the absolute values depend on the methods, they share the magnitude relation; the $t'/t$ value of $\kappa$-(BEDT-TTF)$_2$Cu$_2$(CN)$_3$ is close to unity and decreases in the order $\kappa$-(BEDT-TTF)$_2$Cu$_2$(CN)$_3$ $>$ $\kappa$-(BEDT-TTF)$_2$Cu(NCS)$_2$ $>$ $\kappa$-(BEDT-TTF)$_2$Cu[N(CN)$_2$]Cl $>$ $\kappa$-(BEDT-TTF)$_2$Cu[N(CN)$_2$]Br. Thus, $\kappa$-(BEDT-TTF)$_2$Cu$_2$(CN)$_3$ is closest to the right triangular lattice system among the half-filled compounds. Because the critical $U/W$ value of the Mott transition depends on the $t'/t$ value \cite{mizusaki2006gapless, kyung2006mott, acheche2016mott, szasz2021phase}, it is difficult to precisely determine the critical value by comparing compounds with considerably different values of $t'/t$. As far as the isostructural compounds with relatively close values of $t'/t$ are concerned, the $U/W$ value is larger for the Mott insulating compound $\kappa$-(BEDT-TTF)$_2$Cu[N(CN)$_2$]Cl than for the metallic compound $\kappa$-(BEDT-TTF)$_2$Cu[N(CN)$_2$]Br, which is consistent with the bandwidth-controlled Mott transition \cite{kanoda1997electron}. 

Molecules are mainly held together by van der Waals bonds or weak ionic bonds in organic crystals, whereas inorganic crystals consist of stronger chemical bonds between atoms. The difference in the strength of chemical bonds leads to a compressibility for organic crystals that is higher than that for inorganic crystals. For example, the compressibility values are 6.8\%/GPa for $\kappa$-(BEDT-TTF)$_2$Cu(NCS)$_2$ \cite{rahal1997isothermal} and 1.5\%/GPa for a high-$T_{\rm c}$ cuprate HgBa$_2$CuO$_{4+\delta}$ \cite{hyatt2001structure}. The soft lattice of organic crystals makes bandwidth control by pressure easier than that of inorganic materials, as shown in the band structure calculations for the structural data of $\kappa$-(BEDT-TTF)$_2$Cu(NCS)$_2$ under pressure. This bandwidth controllability enables the examinations of correlated electron physics with pressure experiments \cite{kanoda1997electron}. Indeed, the Mott insulators, $\kappa$-(BEDT-TTF)$_2$Cu[N(CN)$_2$]Cl and $\kappa$-(BEDT-TTF)$_2$Cu2(CN)$_3$, undergo a bandwidth-controlled Mott transition at moderate pressures of 30 MPa \cite{kanoda1997electron,kagawa2004transport} and 150 MPa \cite{geiser1991superconductivity,kurosaki2005mott,furukawa2018quasi}, respectively.

When the band filling deviates from half by carrier doping, the doped carriers can be conductive even under the prohibition of double occupancy. Although the doped carriers tend to be trapped by polaron formation, Anderson localization, or charge ordering, they can be conductive as in high-$T_{\rm c}$ cuprates. Then, what will happen when $U/W$ is varied in such doped systems? The system will always be conductive, but prohibited double occupancy is allowed with decreasing $U/W$. In Chapter 4, an implication of the “bandwidth-controlled Mottness transition” in such situations is discussed with the experimental studies of $\kappa$-Hg$_{2.89}$Br$_8$.

\subsection{Quantum spin liquid}
In the Mott insulating state of $\kappa$-type BEDT-TTF compounds, an antibonding HOMO is an open shell, and a localized electron hosts a spin degree of freedom. A finite probability of inter-dimer electron transfer causes a magnetic interaction between spins, which is usually antiferromagnetic. Although classical spins are expected to form AF order, quantum mechanical spins can form a spin singlet, and thus the magnetism of Mott insulators is nontrivial \cite{lee2008end}. The relative energy between the AF order and spin singlet depends on the lattice dimension; a spin-singlet state has lower energy in a one-dimensional system (Fig. 4(a)). In a square lattice, whose ground state is an AF order, a singlet state has a closer energy to it than in a cubic lattice (Fig. 4(b,c)). In addition to dimensionality, antiferromagnetically interacting spins on a triangular lattice are a typical example of geometrical frustration, as depicted in Fig. 4(d). Thus, a triangular lattice is of great interest because geometrical frustration makes the spin states more versatile as discussed in several review papers \cite{balents2010spin,powell2011quantum,savary2016quantum,zhou2017quantum,pustogow2022thirty}.

%%%%%%%%%%%%%%%%%%%Fig4
\begin{figure}
\includegraphics[width=86mm]{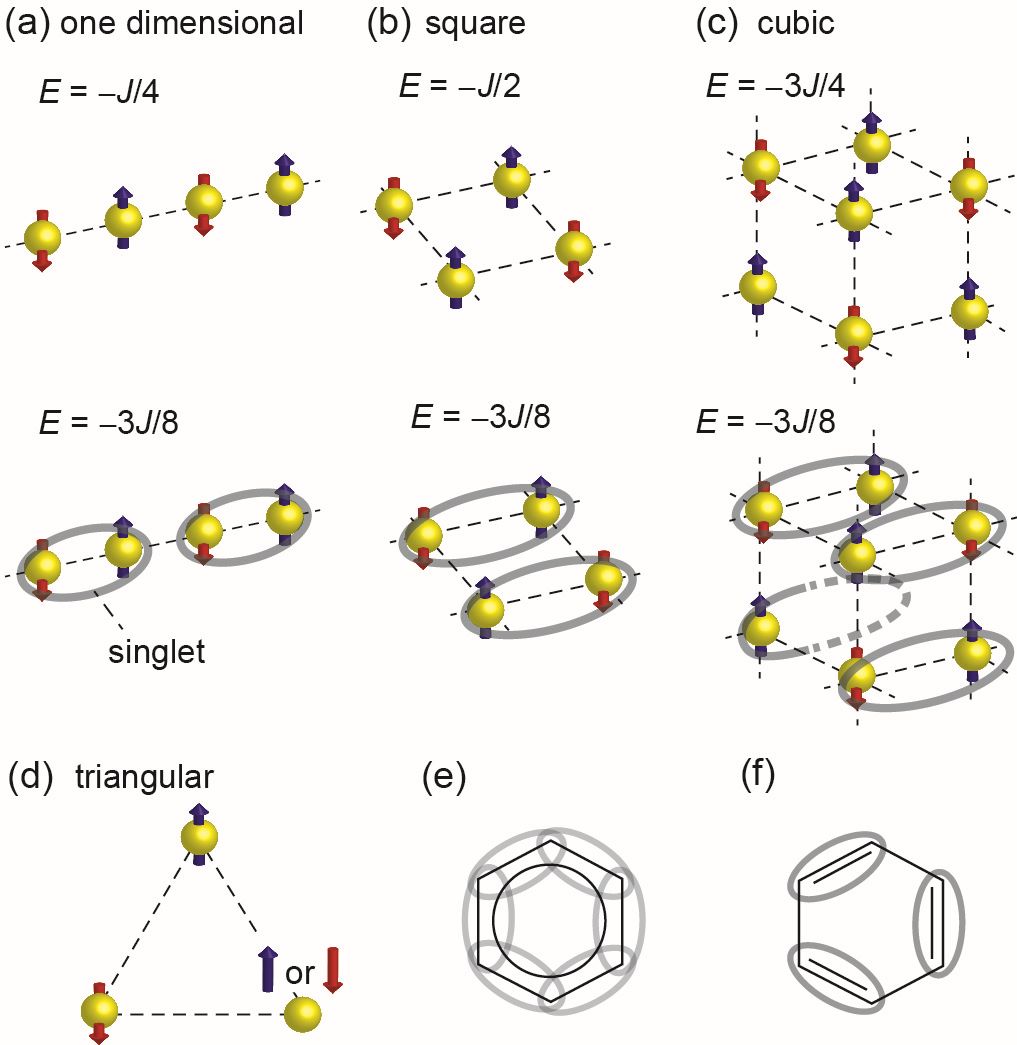}% Here is how to import EPS art
\caption{\label{fig}
(a-c) Antiferromagnetic (AF) order and singlet states in a one-dimensional lattice (a), square lattice (b) and cubic lattice (c). Their energies per site $E$ are calculated with the Heisenberg Hamiltonian ($E = 1/N\Sigma Js\cdot s$) of spin $s$ one half, where $N$ and $J$ represent the number of sites and antiferromagnetic interaction ($J>0$), respectively. (d) Schematic figure of spin frustration on a triangular lattice. (e) A resonating valence bond state of the benzene ring. (f) A hypothetical bond alternating state of the benzene ring. }
\end{figure}
%%%%%%%%%%%%%%%%%%%Fig4

A Mott insulator with a nearly right triangular lattice $\kappa$-(BEDT-TTF)$_2$Cu$_2$(CN)$_3$ is reported to show a QSL-like properties, whereas a Mott insulator with an anisotropic triangular lattice $\kappa$-(BEDT-TTF)$_2$Cu[N(CN)$_2$]Cl undergoes AF ordering \cite{miyagawa1995antiferromagnetic}. The QSL state was suggested in the nuclear magnetic resonance (NMR) experiments as an absence of the spectral splitting or broadening characteristic of AF ordering down to 50 mK \cite{shimizu2003spin}. Then, exotic low-energy excitations were examined in terms of a variety of physical quantities, such as the NMR relaxation rate \cite{shimizu2006emergence}, specific heat \cite{yamashita2008thermodynamic}, thermal conductivity \cite{yamashita2009thermal}, lattice expansion \cite{manna2010lattice}, muon spin resonance \cite{pratt2011magnetic}, and electron spin resonance \cite{miksch2021gapped}. We mention, however, that some results appear not to have coherent explanations at present and even a qualitative property, e.g., whether the excitations are gapless or not, is still under debate partially because of its sensitivity to lattice defects or impurities \cite{furukawa2015quantum,urai2019disorder}, which are inevitably implemented in materials. 

From a theoretical point of view, the resonating-valence-bond (RVB) state was proposed by P. W. Anderson as an analogy to the benzene ring \cite{anderson1973resonating}. It is well known that the $\pi$ orbitals of the benzene ring form six equivalent chemical bonds (Fig. 6(e)), not alternating single and double bonds (Fig. 6(f)). Because the double bond corresponds to a spin singlet state of $\pi$ electrons, a benzene ring can be viewed as a superposition of spin singlet states as in the RVB state. Although the RVB state illustrates a key element of the QSL, recent theoretical studies have revealed a possible variation in QSL states such as gapless spinon Fermi surfaces \cite{motrunich2005variational}, spinon pairing \cite{lee2007amperean, galitski2007spin}, Dirac QSL \cite{hu2019dirac} and chiral QSL \cite{szasz2020chiral}. Besides this diversity, randomness is suggested to alter the QSL nature \cite{kawamura2019nature}, and thus, makes a comparison between experiments and theory elusive.

While studies on spin excitations are thus underway, charge excitations in quantum spin liquids are also intriguing. The charge excitations of $\kappa$-(BEDT-TTF)$_2$Cu$_2$(CN)$_3$ are gapped at ambient pressure, but the charge gap becomes vanishingly small under pressure toward the critical pressure of bandwidth-controlled Mott transition. Then, charge degrees of freedom would be mixed with spin degrees of freedom \cite{furukawa2018quasi}, possibly forming an even more exotic spin and charge liquid. Another intriguing issue is what occurs when carriers are doped into a QSL. The doped RVB state is also proposed by P. W. Anderson as a metallic state, hosting high-$T_{\rm c}$ superconductivity \cite{anderson1987resonating}. However, it is not straightforward to make experimental access to this issue because the mother materials of doped Mott insulators have the AF ground state in most cases. The family of $\kappa$-type BEDT-TTF compounds is unique in that it contains both non-doped and doped triangular lattice compounds, $\kappa$-(BEDT-TTF)$_2$Cu$_2$(CN)$_3$ and $\kappa$-Hg$_{2.89}$Br$_8$. In Chapter 5, the unconventional excitations of doped spin liquid are discussed in detail by comparing the experimental data of these two compounds. 

\subsection{Superconductivity and energetic hierarchy}
Quantum-mechanically degenerate degrees of freedom are often unstable at low temperatures, and can form an ordered state by lifting the degeneracy. For instance, a density wave is formed via Fermi surface nesting and a superconducting state is formed via Cooper pairing. In correlated electron materials, the lifting of degeneracy relates to multiple energy scales, and we refer to such coexistence of energy scales as energetic hierarchy. 

What does it look like to trace the energetic hierarchy of $\kappa$-(BEDT-TTF)$_2$Cu[N(CN)$_2$]Cl from high energy to low energy, or from high temperature to low temperature? Concerning the conduction electrons in the antibonding HOMO band, the largest energy scales are the onsite Coulomb repulsion $U$ and the bandwidth $W$, which are on the order of 10$^3$-10$^4$ K, as schematically summarized in Fig. 5. According to the scaling analysis of resistivity \cite{furukawa2015quantumcri}, the electron system is in a quantum critical regime in the range of a few hundred Kelvin. This implies that the electronic system is neither insulating nor metallic because of the competing nature of $U$ and $W$ and that an insulating phase formed by particle-like electrons and a metallic phase formed by wave-like electrons are quantum-mechanically degenerate. In a temperature range below 100 K, the pressure dependence of the electronic state becomes clear, indicating that the degeneracy of insulating and metallic phases is lifted.

%%%%%%%%%%%%%%%%%%%Fig5
\begin{figure}
\includegraphics[width=86mm]{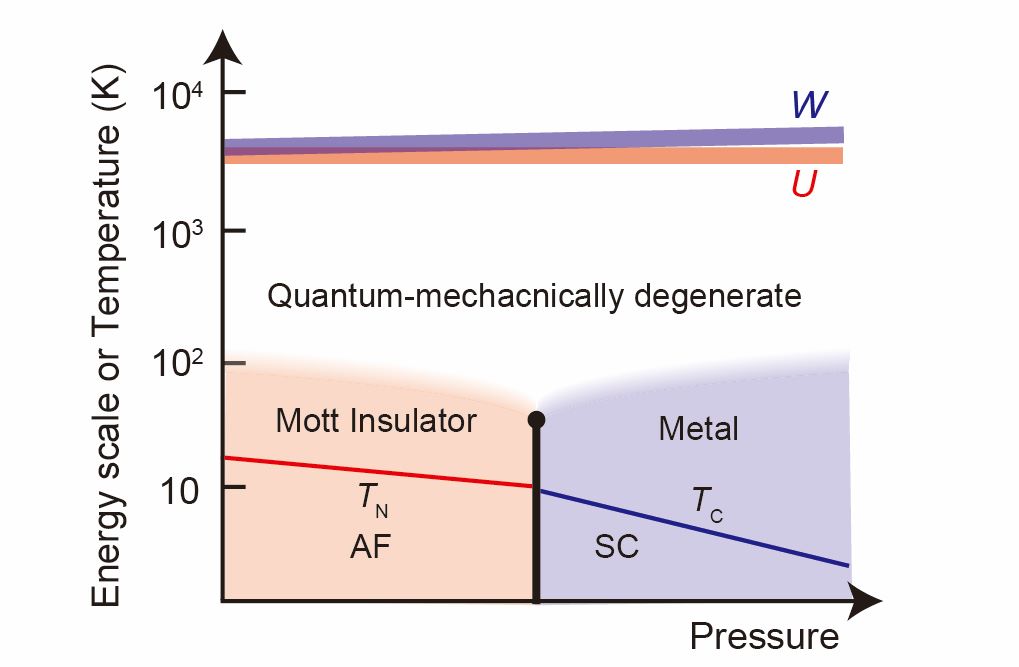}
\caption{\label{fig}
Schematic pressure-temperature phase diagram of $\kappa$-(BEDT-TTF)$_2$Cu[N(CN)$_2$]Cl. Because the energy scales of electronic correlation $U$ and bandwidth $W$ are one order of magnitude larger than that of thermal fluctuations at room temperature, electrons in the conduction band are neither delocalized as waves nor localized as particles. In this sense, electronic states are quantum mechanically degenerate as shown in the scaling analysis of resistivity \cite{furukawa2015quantumcri}. In the temperature range of a few tens of Kelvin, Mott insulating and metallic states are separated by a first-order phase transition represented by the thick black line \cite{kagawa2004transport}. The Mott insulating state undergoes an antiferromagnetic (AF) transition at $T_{\rm N}$, and the metallic state undergoes a superconducting (SC) transition at $T_{\rm c}$. }
\end{figure}
%%%%%%%%%%%%%%%%%%%Fig5

In the low-pressure Mott-insulating regime, further cooling results in magnetic ordering at a temperature of a few tens of Kelvin in $\kappa$-(BEDT-TTF)$_2$Cu[N(CN)$_2$]Cl  \cite{kanoda1997electron,kanoda1997hyperfine}, indicating that the degeneracy of spin degrees of freedom is also lifted. On the other hand, in the high-pressure metallic regime, the metallic phase undergoes a superconducting transition by lifting the Fermi degeneracy and forming a superconducting gap. It should be noted that the energy scales of magnetic ordering or the superconducting gap are lower than the bare interaction strength $U$ and $W$ by more than two orders of magnitude. Thus, to clarify how superconductivity emerges from the quantum mechanical degeneracy, it is necessary to properly trace the energetic hierarchy.

In a doped Mott insulator, the doped carriers can be conductive, even in the presence of a large $U$ value exceeding $W$. Then, what does the energetic hierarchy look like when a doped Mott insulator exhibits superconductivity? A good starting point to trace the energetic hierarchy may not be a conventional Fermi liquid in which electron correlation is regarded as a perturbation. This review traces the energetic hierarchy of $\kappa$-Hg$_{2.89}$Br$_8$ from high energy to low energy, namely, the synthesis of the crystal in Chapter 3, Mott physics in Chapter 4, magnetism in Chapter 5, and superconductivity in Chapter 6.

\section{Synthesis and crystal structure of the doped organic conductor}
Although $\kappa$-(BEDT-TTF)$_4$Hg$_{2.89}$Br$_8$ ($\kappa$-Hg$_{2.89}$Br$_8$) has an almost identical molecular arrangement to the half-filled $\kappa$-type BEDT-TTF compounds mentioned in Chapter 2, it hosts two sublattices that are incommensurate with each other \cite{lyubovskaya1987superconductivity, li1998room}. One sublattice is composed of BEDT-TTF molecules and Br ions, and the other is formed by Hg ions. Although it is not clarified why the chemistry of BEDT-TTF, Br, and Hg self-organizes such an incommensurate structure, the band filling of $\kappa$-Hg$_{2.89}$Br$_8$ deviates from half owing to this non-stoichiometry. $\kappa$-Hg$_{2.89}$Br$_8$ is a unique material in that it is superconducting at low temperatures despite being a doped organic material. In this chapter, synthetic methods and crystal structures are described to consider what is currently known about the chemistry and metallicity of $\kappa$-Hg$_{2.89}$Br$_8$.

\subsection{Synthesis}
Most BEDT-TTF salts are synthesized by electrolytic oxidation of the solution in which BEDT-TTF molecules and supporting electrolyte are dissolved. Oxidized BEDT-TTF becomes a radical and binds to the cationic electrolyte. In the earliest synthesis methods, the electrolytes were relatively simple tetrabutylammonium (TBA) salts (e.g., TBAI$_3$, TBAIBr$_2$ in the cases of the anions of I$_3$ and IBr$_2$, respectively) \cite{saito1982two,kobayashi1983transverse,parkin1983superconductivity,yagubskii1984normal,hennig1985alpha,kobayashi1986crystalicl2,anzai1987crystal}. 

Later, the development of synthesis methods enabled to crystalize the compounds whose counter anions are present only in solution, not in an ingredient salt. For instance, in $\kappa$-(BEDT-TTF)$_2$Cu(NCS)$_2$ \cite{urayama1988new,urayama1988crystal}, the counter anion [Cu(NCS)$_2$]$^-$ is not present as an isolated ion in its ingredient salts CuSCN and KSCN. Most $\kappa$-type BEDT-TTF compounds are synthesized with this method, and this is also the case for $\kappa$-Hg$_{2.89}$Br$_8$, where the divalent anion [Hg$_3$Br$_8$]$^{2-}$ is required for synthesis \cite{lyubovskaya1991controlled}. [Hg$_3$Br$_8$]$^{2-}$ exists only as an equilibrium state in a solution, as shown in the following equation: 
\\
\\
2[HgBr$_3$]$^-$ + n[HgBr$_2$] $\leftrightarrow$ [Hg$_3$Br$_8$]$^{2-}$.
\\
\\
Therefore, it is necessary to first prepare this divalent anion for the synthesis of $\kappa$-Hg$_{2.89}$Br$_8$ from the ingredients TBAHgBr$_3$ and  HgBr$_2$. The density of [Hg$_3$Br$_8$]$^{2-}$ can be adjusted by the amount of HgBr$_2$ added to the solvent. In addition to the preparation of [Hg$_3$Br$_8$]$^{2-}$, the subsequent electrochemical synthesis is complex. If the amount of HgBr$_2$ is too high, then nucleation of a crystal frequently occurs during electrochemical synthesis, resulating in a large number of small crystals. To obtain a large single crystal, we followed the synthesis conditions described in the literature \cite{lyubovskaya1991controlled} as much as possible by fully considering the above factors.

In our synthesis, we first placed two kinds of solutes, TBANHgBr$_3$ and HgBr$_2$, in a 1,1,2-trichloroethane solvent with concentrations of 18.7 mmol/L and 0.30 mmol/L, respectively, and stirred for one day. Since previous studies \cite{lyubovskaya1991controlled} highlight importance of the purity of the former solute, it was purified twice by a recrystallization method. After confirming that all solutes had dissolved, BEDT-TTF molecules were added to the solution at a concentration of 1.6 mmol/L and stirred again for another day. The solution was poured into an electrochemical glass cell and left in an incubator at 40 °C overnight. Then, electrochemical oxidation was started by applying a current of 0.5 $\mu$A. Even if we follow the above procedure, $\kappa$-Hg$_{2.89}$Br$_8$ often happens not to grow due to polymorphism. In our experience, an electrochemical synthesis with a relatively long period of time (2 or 3 months) enhances the probability of obtaining $\kappa$-Hg$_{2.89}$Br$_8$. Therefore, we use large glass cells (e.g., 100 mL or 200 mL) for long-term electrochemical synthesis.

Crystals of $\kappa$-Hg$_{2.89}$Br$_8$ are easily distinguished from other polymorphs by their shapes; $\kappa$-Hg$_{2.89}$Br$_8$ has a rhombic shape as shown in Fig. 6(a), whereas other products are needle-shaped or rectangular crystals. However, we occasionally encountered crystals whose shapes are indistinguishable. In such a case, electrical resistivity measurements at room temperature enable the distinction because the other products have much higher electrical resistivity than $\kappa$-Hg$_{2.89}$Br$_8$. Thus, the complex synthesis method of $\kappa$-Hg$_{2.89}$Br$_8$ has been established.

%%%%%%%%%%%%%%%%%%%Fig6
\begin{figure}
\includegraphics[width=86mm]{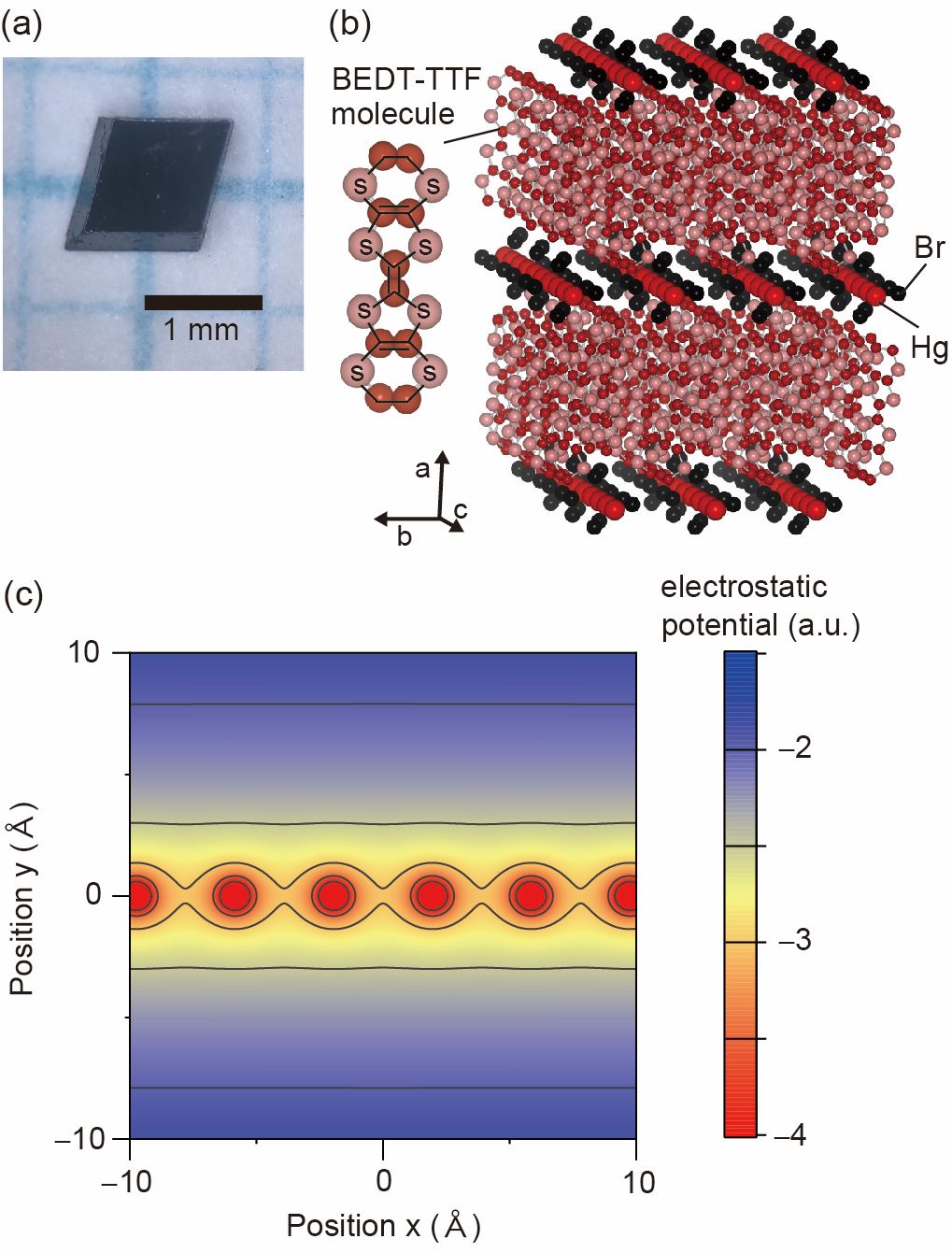}
\caption{\label{fig}
(a) Picture of a single crystal of $\kappa$-(BEDT-TTF)$_4$Hg$_{2.89}$Br$_8$. (b) Crystal structure of $\kappa$-(BEDT-TTF)$_4$Hg$_{2.89}$Br$_8$. (c) Calculated electrostatic potential generated by Hg ions. One hundred Hg ions are one-dimensionally aligned along the x-axis with an interval of 3.9 \AA \ at $y = 0$, and the electrostatic potential from each ion is superimposed. At distances greater than 3 \AA \ from the Hg ions, the x dependence of the potential is less pronounced, and the conduction electrons are less affected by the incommensurability of the electrostatic potential. }
\end{figure}
%%%%%%%%%%%%%%%%%%%Fig6

\subsection{Crystal structure}
 $\kappa$-Hg$_{2.89}$Br$_8$ has a unique crystal structure, where BEDT-TTF molecules are commensurate with Br ions but incommensurate with Hg ions (Fig.6(b))  \cite{lyubovskaya1987superconductivity, li1998room}. Because the counter anion [Hg$_3$Br$_8$]$^{2-}$ is stoichiometric, a few percent of the large Hg ions seem to be expelled from the crystal during the crystallization process, possibly due to strain release. Although the mechanism is not fully clarified, the chemistry of BEDT-TTF, Br, and Hg results in the fixed composition of (BEDT-TTF)$_4$Hg$_{2.89}$Br$_8$. When we assume that the valences of Br and Hg ions are -1 and +2, respectively, the valence of BEDT-TTF is +0.555, which is 11\% larger than that of the $\kappa$-type BEDT-TTF compounds with a half-filled band. The inter-Hg distance of $\kappa$-Hg$_{2.89}$Br$_8$, 3.9 \AA, is much larger than that in Hg$_2$X$_2$ (X = Cl, Br, I) crystals, 2.4 \AA \cite{havighurst1926parameters}, which have inter-Hg bonds; this supports the assumption for the valence of Hg$^{2+}$ with a closed shell. Indeed, Raman spectroscopy revealed the valence of BEDT-TTF by observing valence-sensitive molecular vibration modes \cite{yamamoto2005examination}. According to the calculations based on extended H\"{u}ckel and tight-binding approximations, the onsite Coulomb repulsion $U$ is larger than the bandwidth $W$ (Fig. 3(d)). Thus, $\kappa$-Hg$_{2.89}$Br$_8$ can be viewed as a doped Mott insulator.

When one tries to control band filling, mixed crystals are often used in high-$T_{\rm c}$ cuprates La$_{\rm 2-x}$Sr$_2$CuO$_4$ where La and Sr ions are trivalent and divalent, respectively \cite{imada1998metal}. In molecular crystals, such chemical doping has also been extensively studied, and has achieved metallic states in $\lambda$-(BEDT-TTF)$_2$(GaCl$_4^{-1}$)$_{1-x}$(CoCl$_4^{-2}$)$_x$, $\alpha$-(BEDT-TTF)$_3$(GaCl$_4^{-1}$)$_{1-x}$(CoCl$_4^{-2}$)$_x$(TCE) and $\beta'$-(BEDT-TTF)$_3$(GaCl$_4^{-1}$)$_{2-x}$(CoCl$_4^{-2}$)$_x$ \cite{mori2002first, mori2002control, katsuhara2002band, suto2005metallic}. However, superconductivity has not been achieved in these organic mixed crystals as they exhibit an nonmetallic temperature dependence of resistivity at low temperatures. One possible reason for the nonmetallic behaviors is the localization of carriers due to an irregular electrostatic potential from randomly distributed multiple kinds of anions. The disorder effect is pronounced in organic materials, whose bandwidth is typically several to ten times smaller than that of inorganic materials such as high-$T_c$ cuprates. 

In $\kappa$-Hg$_{2.89}$Br$_8$, chemical doping is realized due to the incommensurate crystal structure, which is also the case for $\kappa$-(BEDT-TTF)$_4$Hg$_{2.78}$Cl$_8$ \cite{lyubovskaya1985superconductivity,lyubovskaya1991controlled}, (MDT-TS)(AuI$_2$)$_{0.441}$ \cite{kawamoto2005superconductivity} and (MDT-ST)(I$_3$)$_{0.417}$ \cite{kawamoto2006fermi}. The Hg lattice is incommensurate with the BEDT-TTF lattice along the c-axis, which is parallel to a conducting plane, and commensurate along the a-axis and b-axis, thus producing an incommensurate electrostatic potential along the c-axis. The theoretical study expect the appearance of an incommensurate Mott insulator when the potential corrugation is large enough \cite{seo2008mott}. However, the amplitude of the potential corrugation is vanishingly small in the area where the distance from the Hg ions is sufficiently longer than spacing between Hg ions, 3.9 \AA (Fig. 6(c)). Therefore, the incommensurability of the potential should not significantly affect the conduction electrons because they mainly exist in the TTF structure, the central part of BEDT-TTF molecules, which is approximately 6-9 \AA away from Hg ions. In reality, the incommensurate Hg ions slightly modulate the positions of BEDT-TTF molecules \cite{li1998room}, which might generate an irregular potential to conduction electrons as a higher order effect. Thus, the irregular potential, which inevitably accompanies chemical doping, affects electrical conduction in a different manner from mixed crystals. This may explain why $\kappa$-Hg$_{2.89}$Br$_8$ is metallic at the lowest temperatures and becomes superconducting.

Another characteristic feature in the structure of $\kappa$-Hg$_{2.89}$Br$_8$ is a nearly isotropic triangular lattice \cite{li1998room,oike2015pressure,oike2017anomalous}. According to the band structure calculations \cite{oike2015pressure,oike2017anomalous}, the value of $t'/t$, which indicates an anisotropy of the triangular lattice, is close to that of the QSL candidate material $\kappa$-(BEDT-TTF)$_2$Cu$_2$(CN)$_3$. This feature makes the electronic states of $\kappa$-Hg$_{2.89}$Br$_8$ even more intriguing because it enables us to examine the behaviors of correlated electrons in a doped Mott insulator with a triangular lattice as well as the impact of pressure variation on them.

\section{Mottness in the doped organic conductor}
When the on-site Coulomb repulsion $U$ is sufficiently larger than the bandwidth $W$, double occupancy of electrons at a lattice site is strongly prohibited. The tendency or degree of prohibited double occupation is often referred to as Mottness \cite{phillips2006mottness}. In a half-filled system, where the number of electrons is the same as that of lattice sites, Mottness drives the electronic system into a Mott insulating state \cite{kotliar2004strongly}. However, off the half filling, electrons can be conductive even if double occupancy is prohibited, and thus Mottness does not necessarily cause an insulating state \cite{capone2004strongly,sordi2010finite,sordi2012strong,sordi2012pseudogap,yokoyama2012crossover,gull2013superconductivity,simkovic2022origin}. Thus, how Mottness matters in doped systems is an intriguing issue. This chapter describes the electrical conduction phenomena exhibited by the doped system, based on the experimental results of resistivity and Hall measurements under pressure, which varies Mottness.

\subsection{Charge transport under pressure}
Four-probe resistivity measurement is a straightforward approach to study electrical conduction phenomena. However, the inhomogeneity of local resistivity in a sample may obscure the interpretation of measurement results. The current path is determined so that the resistance is minimized and thus may also be inhomogeneous. The measurements under pressure, which may induce inhomogeneity in the internal pressure, should be even more cautious about this. In this respect, liquids are employed as pressure media because the fluidity of the liquid homogenizes the pressure at the sample surface. This allows isotropic pressure, or hydrostatic pressure, to be applied up to the solidification pressure of the pressure medium.

%%%%%%%%%%%%%%%%%%%Fig7
\begin{figure}
\includegraphics[width=86mm]{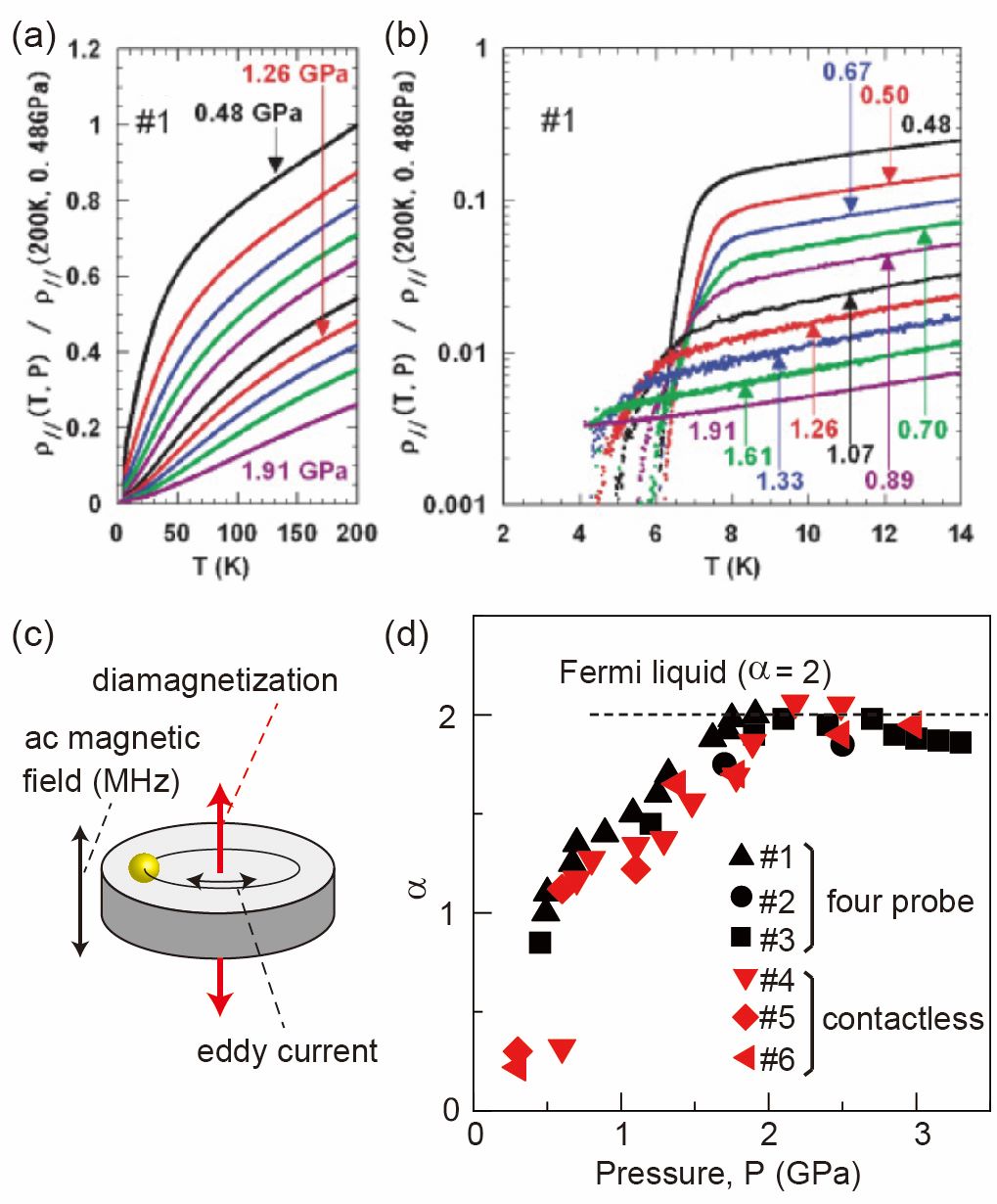}
\caption{\label{fig}
(a) Temperature dependence of the in-plane resistivity $\rho_{//}$ of $\kappa$-(BEDT-TTF)$_4$Hg$_{2.89}$Br$_8$ under pressure \cite{taniguchi2007anomalous}. Reprinted from Ref. \cite{taniguchi2007anomalous}. \copyright{2007 The Physical Society of Japan}. (b) Low temperature expansion of the temperature dependences of $\rho_{//}$ \cite{taniguchi2007anomalous}. Reprinted from Ref. \cite{taniguchi2007anomalous}. \copyright{2007 The Physical Society of Japan}. (c) Schematic figure of the contactless conductivity measurement. When an ac field is applied to a metallic material, diamagnetization appears due to electromagnetic induction. (d) Pressure dependence of the exponent $\alpha$ in $\rho_{//}$ = $\rho_{o} + AT^{\alpha}$, which fits the data in a temperature range below 20 K and above the superconducting transition temperatures. $\rho_{o}$ and $A$ are the residual resistivity and the temperature coefficient, respectively \cite{taniguchi2007anomalous, oike2015pressure}. }
\end{figure}
%%%%%%%%%%%%%%%%%%%Fig7

%%%%%%%%%%%%%%%%%%%Fig8_wide
\begin{figure*}
\includegraphics[width=172mm]{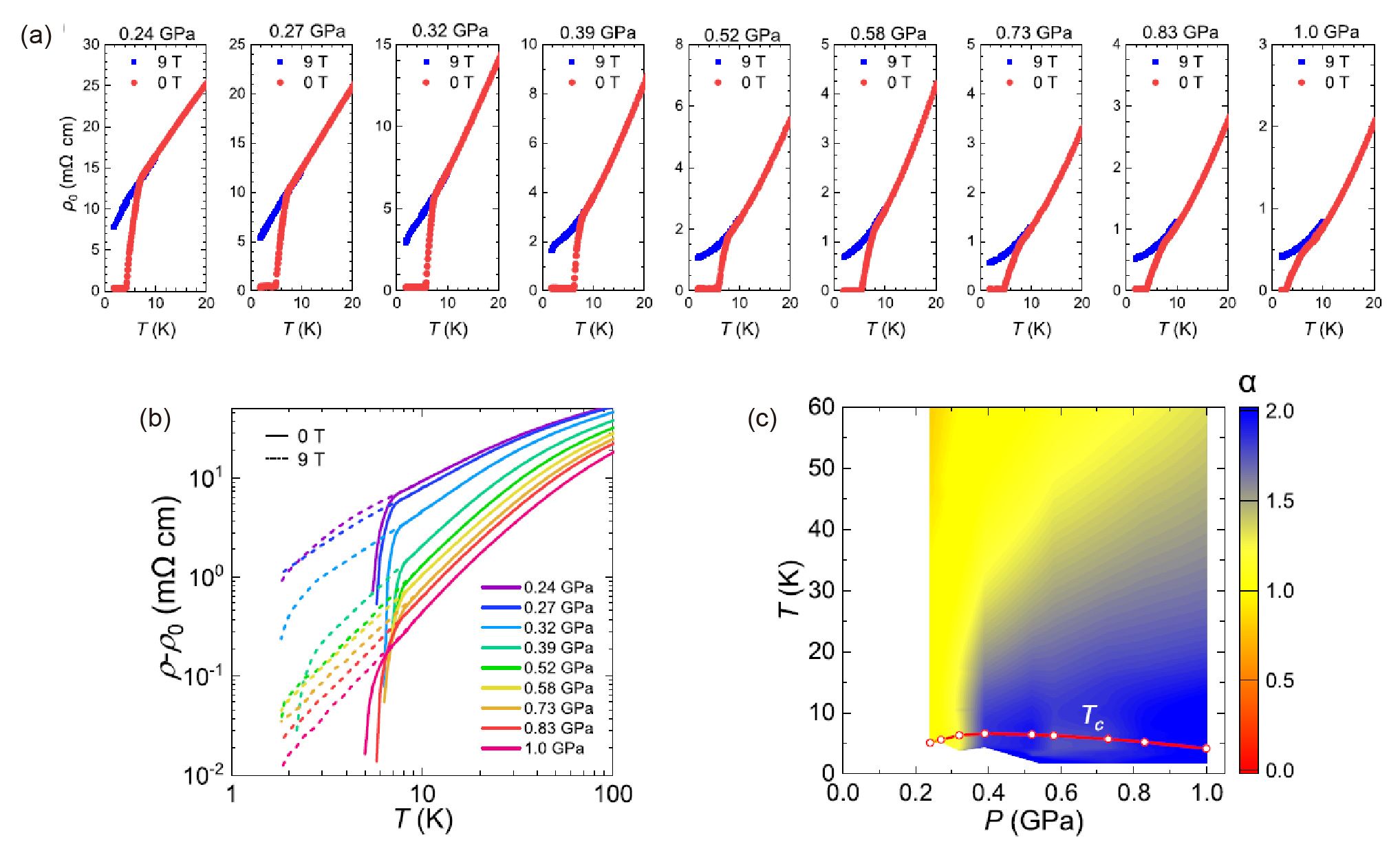}
\caption{\label{fig:wide}
(a) Temperature dependence of the in-plane resistivity $\rho$ of $\kappa$-(BEDT-TTF)$_4$Hg$_{2.89}$Br$_8$ at 0 T and a perpendicular field of 9 T under pressure. Reprented from Ref. \cite{suzuki2022mott} under the creative commons license. (b) Log-log plot of $\rho$($T$)$-$$\rho_{o}$. $\alpha$ is determined by $\alpha$ $\equiv$ $d$log[$\rho$($T$)$-$$\rho_{o}$]/$d$log$T$, which corresponds to the slope in the log-log plot. Reprented from Ref. \cite{suzuki2022mott} under the creative commons license. (c) Contour plot of pressure-temperature dependence of the exponent $\alpha$. The yellow region ($\alpha \approx 1$) corresponds to a non-Fermi liquid phase, and the blue region ($\alpha \approx 2$) corresponds to a Fermi liquid phase. Reprented from Ref. \cite{suzuki2022mott} under the creative commons license.}
\end{figure*}
%%%%%%%%%%%%%%%%%%%Fig8_wide

The first transport study on $\kappa$-Hg$_{2.89}$Br$_8$ under pressure was reported by Bud’ko et al. in 1992 \cite{bud1992anomalous}. According to the paper, the electrical resistivity along the conduction plane $\rho_{//}$ is metallic in a low-pressure range, whereas nonmetallic behaviors appear in a high-pressure range above 3.0 GPa. It is noted, however, that most pressure media solidify before reaching 3.0 GPa during pressure application at room temperature. More than a decade later, a pressure medium Daphne 7373, which does not solidify up to 1.9 GPa \cite{yokogawa2007solidification}, was developed and used for resistivity measurements in the high-pressure range of $\kappa$-Hg$_{2.89}$Br$_8$. Then, it was found that nonmetallic behaviors do not appear and that a metallic phase with $\rho_{//}$ obeying the temperature($T$)-square law dominates in a pressure range of 1.5-3.4 GPa (Fig. 7 (a,b)) \cite{okuhata2007high, taniguchi2007anomalous}. Thus, the nonmetallic behaviors in the earlier report possibly originate from the anisotropy and/or inhomogeneity of the pressure. Indeed, we confirmed that intentionally introduced uniaxial strain drives $\kappa$-Hg$_{2.89}$Br$_8$ into a nonmetal \cite{oike2022metal}. The quality of pressure occasionally plays a decisive role in the electronic state as in the present case. 

A way to minimize inhomogeneity in current path, which inevitably depends on the geometry of electrodes attached to the crystal, is to perform contactless conductivity measurements, which are ac susceptibility measurements in the MHz frequency range. When an ac magnetic field is applied to a metallic material, eddy currents induced by electromagnetic induction cause a diamagnetic response (Fig. 7(c)). As a result, the magnetic flux penetrates the sample from the surface only by the skin depth, which is analogous to the penetration depth in a superconducting phase. This skin effect enables probing resistivity via an ac susceptibility measurement with a different geometry of the current path from the conventional four-probe method because the skin depth depends on resistivity \cite{oike2015pressure,oike2009contactless}. The contactless measurements were performed with the pressure medium Daphne 7474, which does not solidify up to 3.7 GPa \cite{murata2008pressure}. Because Daphne 7474 degrades electrodes made of carbon paste, which is often used in the transport measurements of organic conductors, the contactless method is also advantageous in terms of avoiding contact resistance problems. In all four-probe measurements with Daphne 7373 or a cubic anvil cell and contactless measurements with Daphne 7474, $\rho_{//}$ at low temperatures follows the $T$-square law of Fermi liquid in a high-pressure range, whereas it deviates from the $T$-square law in a low-pressure range (Fig. 7(d)) \cite{taniguchi2007anomalous,oike2015pressure}. The consistent results obtained by the different methods suggest that the emergence of the Fermi liquid at high pressures is due to the bulk nature of $\kappa$-Hg$_{2.89}$Br$_8$.

The temperature-pressure dependence of the exponent $\alpha$ in $\rho_{//}$ = $\rho_{o} + AT^{\alpha}$ is essential for understanding how electron correlations affect their conduction, where $\rho_{o}$ and $A$ are the residual resistivity and the temperature coefficient, respectively. However, the emergence of superconductivity in a pressure range below 1.5 GPa hinders the determination of $\rho_{o}$, which is necessary to obtain an accurate value of $\alpha$. Then, the temperature-pressure dependence was investigated by suppressing superconductivity with a magnetic field to determine the $\rho_{o}$ values \cite{suzuki2022mott}. These experiments revealed that the $\rho_{//}$ behaviors at low temperatures exhibit an abrupt change from $T$ linear ($\alpha$ = 1) to $T$ square ($\alpha$ = 2) at a critical pressure of approximatly 0.4 GPa (Fig. 8)  \cite{suzuki2022mott}. 

%%%%%%%%%%%%%%%%%%%Fig9
\begin{figure}
\includegraphics[width=86mm]{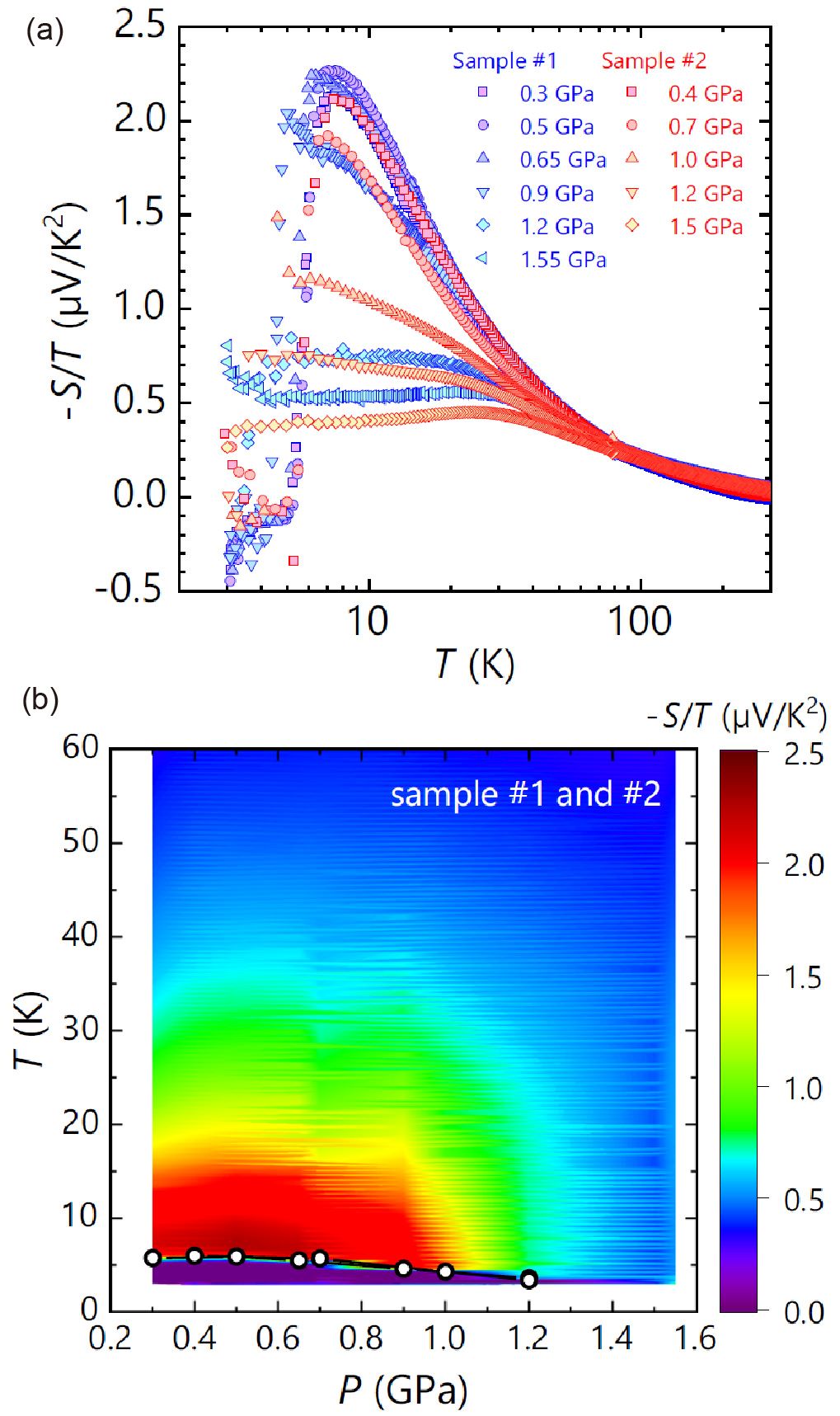}
\caption{\label{fig:wide}
(a) Temperature dependence of the seebeck coeficient divided by temperature $S$/$T$ of $\kappa$-(BEDT-TTF)$_4$Hg$_{2.89}$Br$_8$ under pressure. (b) Contour plot of pressure-temperature dependence of $S$/$T$. Reprented from Ref. \cite{wakamatsu2023thermoelectric} under the creative commons license.}
\end{figure}
%%%%%%%%%%%%%%%%%%%Fig9

A $T$-linear behavior of resistivity is also observed in the doped Mott insulators, high-$T_{\rm c}$ cuprates \cite{imada1998metal} but not in the non-doped $\kappa$-type BEDT-TTF compounds \cite{furukawa2018quasi}. This comparison suggests that doped carriers in a Mott insulator are likely to be responsible for the anomalous metallicity in $\kappa$-Hg$_{2.89}$Br$_8$. Although the mechanism for $T$-linear behavior has not been identified, we discuss how to view anomalous metallicity. One view is that some fluctuations relevant to electron transport diverge toward lower temperatures, as in the case of the quantum critical point in heavy fermion systems \cite{stewart2001non}. When a power-law divergence toward absolute zero is superimposed, the resultant temperature dependence deviates from the $T$-square behavior of a conventional Fermi liquid \cite{stewart2001non}. In this scenario, $\kappa$-Hg$_{2.89}$Br$_8$ is considered to be in a “quantum critical phase” in a pressure range below 0.4 GPa. A very recent study of the Seebeck effect in $\kappa$-Hg$_{2.89}$Br$_8$ supports this point of view (Fig.9) \cite{wakamatsu2023thermoelectric}; in the low-pressure range where a non-Fermi liquid emerges, the Seebeck coefficient divided by temperature, -$S$/$T$ exhibits logarithmic temperature dependence, which has been observed in a wide variety of materials \cite{daou2009thermopower,laliberte2011fermi,boulanger2020transport,mandal2019anomalous,gooch2009evidence,arsenijevic2013signatures,maiwald2012signatures,hartmann2010thermopower,malone2012thermoelectricity,kuwai2011thermoelectric,mun2010thermoelectric,mun2013magnetic,matusiak2011quantum,limelette2010quantum}  and discussed theoretically as an indication of quantum criticality \cite{paul2001thermoelectric,kim2010thermopower,buhmann2013numerical,georges2021skewed}, while -$S$/$T$ is temperature-insensitive at high pressures as expected in a Fermi liquid. Another view is that charge excitation or charged particles that follow unusual statistics emerge due to Mottness \cite{zaanen2011mottness,yamaji2011composite,patel2017quantum}. Because Mottness imposes constraints on electrical transport due to the suppression of double occupancies, low-energy excitations may appear to be different particles from conventional quasiparticles, possibly resulting in a deviation of the $\alpha$ value from 2. These views are not necessarily incompatible with each other. Clarifying the mechanism responsible for anomalous metallicity is a future challenge because it can correlate the transport phenomena of correlated electrons in inorganic and organic materials.

\subsection{Hall effect and Mottness}
The Hall coefficient $R_{\rm H}$ is inversely proportional to the carrier density in the simplest view. From the data at 10 K, one can see a clear pressure dependence of $R_{\rm H}$ in $\kappa$-Hg$_{2.89}$Br$_8$ (Fig. 10) \cite{oike2015pressure}. Although $R_{\rm H}$ abruptly changes in a low-pressure range, it exhibits a moderate pressure dependence similar to half-filled systems in a pressure range above 0.6 GPa. Given that $\rho_{//}$ obeys the $T$-square law, the behavior of $R_{\rm H}$ in a high-pressure range can be attributed to a Fermi-liquid picture. On the other hand, the enhancement of $R_{\rm H}$ occurs in a pressure range of $T$-linear behavior beyond the framework of a conventional Fermi liquid. At ambient or low pressures, the electronic specific heat coefficient and nuclear magnetic resonance (NMR) relaxation rate are significantly enhanced \cite{naito2005anomalous, eto2010non}, implying that a strong electronic correlation is likely responsible for the anomalous metallicity. Such anomalous metallicity is also observed in doped Mott insulators with electric-double-layer transistor \cite{kawasugi2019non}.

%%%%%%%%%%%%%%%%%%%Fig10
\begin{figure}
\includegraphics[width=86mm]{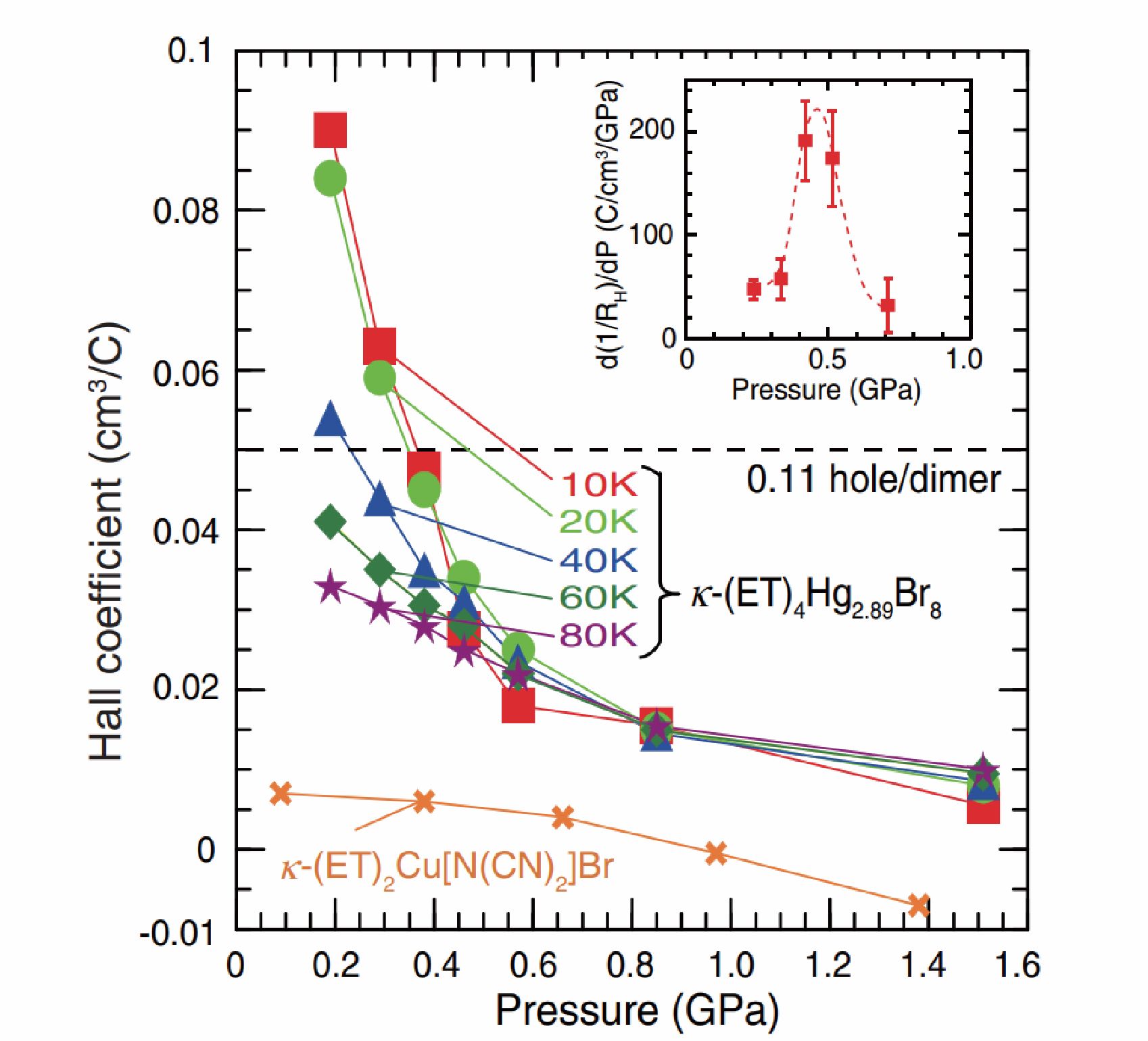}
\caption{\label{fig}
Pressure dependence of the Hall coefficient $R_{\rm H}$ for the doped compound $\kappa$-(BEDT-TTF)$_4$Hg$_{2.89}$Br$_8$ at several temperatures and a half-filled compound $\kappa$-(BEDT-TTF)$_2$Cu[N(CN)$_2$]Br at 10 K. The inset shows the pressure derivative of 1/$R_{\rm H}$ for $\kappa$-(BEDT-TTF)$_4$Hg$_{2.89}$Br$_8$ at 10 K. Reprinted from Ref. \cite{oike2015pressure}. \copyright{2015 the American Physical Society}.}
\end{figure}
%%%%%%%%%%%%%%%%%%%Fig10

%%%%%%%%%%%%%%%%%%%Fig11_wide
\begin{figure*}
\includegraphics[width=172mm]{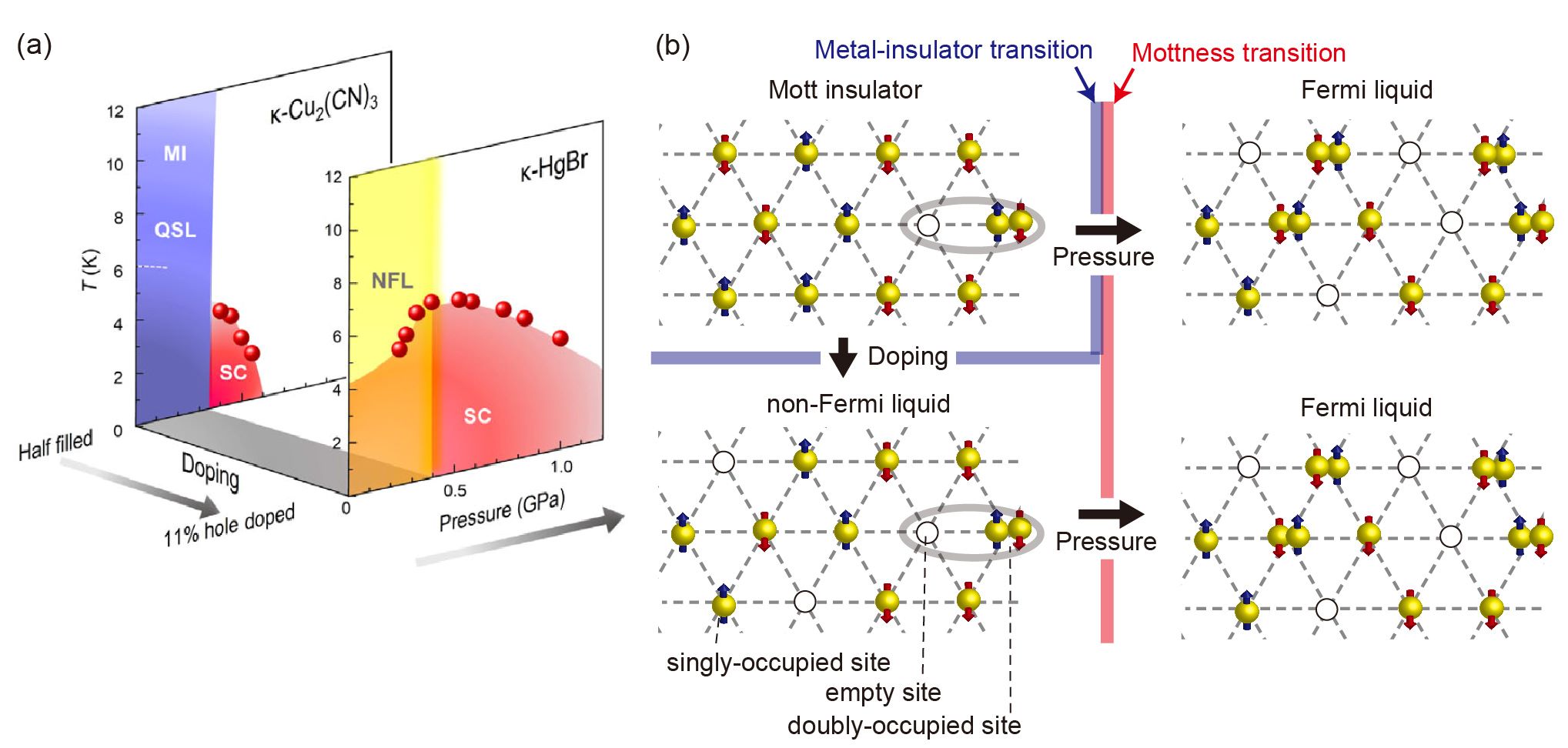}
\caption{\label{fig:wide}
(a) Pressure-temperature phase diagrams of $\kappa$-(BEDT-TTF)$_4$Hg$_{2.89}$Br$_8$ and a quantum spin liquid (QSL) candidate $\kappa$-(BEDT-TTF)$_2$Cu$_2$(CN)$_3$. The red, green and purple areas indicate the superconducting (SC), non-Fermi liquid (NFL), and Mott insulating (MI) phases, respectively. Reprented from Ref. \cite{suzuki2022mott} under the creative commons license. (b) Schematic figure of the metal-insulator transition and Mottness transition. We call the phase transition with respect to Mottness as ``Mottness transition''. A Mottness transition accompanies a metal-insulator transition in half-filled compounds, but does not necessarily accompany it in the doped compound because doped carriers can be mobile even under the prohibition of double occupancy.}
\end{figure*}
%%%%%%%%%%%%%%%%%%%Fig11_wide

Pressure does not change the chemical composition and therefore the band filling of HOMO-based conduction bands is fixed. Thus, it is not straightforward to explain the pressure dependence of $R_{\rm H}$. In general, several factors can contribute to the Hall coefficient, including the carrier density, anisotropy of the carrier mobility, and gauge fields \cite{lange1999magnetotransport,kontani1999hall,haerter2008hall,nagaosa2010anomalous,lee2006doping}. When double occupancy is allowed under pressure, the effective density of mobile carriers increases with pressure. Simply assuming that only doped carriers are mobile, the mobile carrier density $n$ equals 0.11 per site, which corresponds to 5.0 $\times$ 10$^{-2}$ cm$^3$/C in 1/$ne$, where $e$ is the elementary charge. The values below 0.4 GPa are in this range. Another factor is that wavenumber-dependent carrier mobility, which appears, for example, when electrons and holes coexist or when anisotropic scattering is pronounced, affects the $R_{\rm H}$ values as argued in fluctuation-exchange theory and the $t-t_o-J$ model \cite{haerter2008hall}. The spin fluctuations are enhanced due to Mottness, and thus, it is reasonable that applying pressure suppresses this contribution consistently with the reduction in the NMR relaxation rate under pressure \cite{naito2005anomalous,eto2010non}. The third factor is gauge field, which emerges when quantum mechanical particles move under certain constrains \cite{nagaosa2010anomalous}. Recent theoretical studies suggest that anomalous Hall effects emerge in the $\kappa$-type BEDT-TTF compounds with a half-filled band due to interplay between spin structure and band structure \cite{naka2019spin, naka2020anomalous, hayami2020multipole}. In doped Mott insulators, charged carriers are prohibited to doubly occupy a single site, and consequently their wave function can gain a phase factor that depends on their trajectory of constraint motion \cite{lee2006doping}, possibly leading to a kind of anomalous Hall effect. This mechanism is also considered to be suppressed under high pressure together with suppression of Mottness. Although the mechanism responsible for the increase in $R_{\rm H}$ have not been identified, but Mottness lies behind all of them. Therefore, the anomalous metallicity can be seen as a nonperturbative effect of strong electronic correlations on electron transport. 

\subsection{Bandwidth-controlled Mottness transition at a finite doping level}
A comparison of the pressure dependence of $\kappa$-Hg$_{2.89}$Br$_8$with that of the non-doped $\kappa$-type BEDT-TTF compounds reveals a generalized picture of the Mott transition (Fig. 11). In non-doped systems, a drastic change in double-occupancy probability, or a bound-unbound transition of doublons and holons, is associated with the Mott metal-insulator transition, where the doublon and holon stand for doubly occupied and empty sites, respectively \cite{kotliar2004strongly} (Fig. 11(b)). This means a one-to-one correspondence between Mottness and metal-insulator transition. However, when a Mott insulator is doped, the excess electrons and holes can still be conductive, even if double occupancy is prohibited. Then, this correspondence does not hold, and instead, we may refer to the drastic change in double-occupancy probability as ``Mottness transition" (Fig. 11(b)). The $T$-linear behavior of $\rho_{//}$ and the enhancement of $R_{\rm H}$ indicate the separation of the Mottness transition and metal-insulator transition. $\kappa$-Hg$_{2.89}$Br$_8$ is considered to exhibit a bandwidth-controlled Mottness transition at a pressure of approximately 0.5 GPa, above which the anomalous metallicity disappears. 

The critical $U/W$ values of the Mottness transition are estimated to be 1.07 for $\kappa$-Hg$_{2.89}$Br$_8$ and 0.88 for $\kappa$-(BEDT-TTF)$_2$Cu$_2$(CN)$_3$ based on extended H\"{u}ckel and tight binding approximations \cite{oike2015pressure,mori1984intermolecular,geiser1991superconductivity,li1998room,kurosaki2005mott}, implying that carrier doping increases the critical value of $U/W$. The itineracy of doped carriers enhances the screening effect and effectively weakens electron-electron interactions, possibly leading to an increase in the critical value of $U/W$. This picture is in line with the theoretical studies based on dynamical mean field theory (DMFT) \cite{sordi2010finite,sordi2012strong,sordi2012pseudogap,gull2013superconductivity, reymbaut2019pseudogap}, which considers hybridization between an orbital in a lattice site of interest and an electron bath composed of other lattice sites \cite{kotliar2004strongly}. An electronic state at the lattice site is the superposition of empty, single and double occupancy, which self-consistently determines the density of states. Therefore, when the prohibition of double occupancy opens an energy gap in the single-particle excitation at the lattice site, the density of states of the electron bath has an energy gap. DMFT reproduce the phase transitions in the Mottness in both of non-doped and doped systems, and predicts an increase in the critical $U/W$ value by doping \cite{sordi2010finite,sordi2012strong,sordi2012pseudogap,gull2013superconductivity}. Recent theoretical studies have revealed the nature of metallic states that are qualitatively different from Fermi liquids \cite{phillips2020exact,schafer2021tracking,huang2022discrete,simkovic2022origin,qin2022hubbard,vsimkovic2022two}, which should lead to our understanding of anomalous metallicty in doped Mott insulators.

%%%%%%%%%%%%%%%%%%%Fig12_wide
\begin{figure*}
\includegraphics[width=172mm]{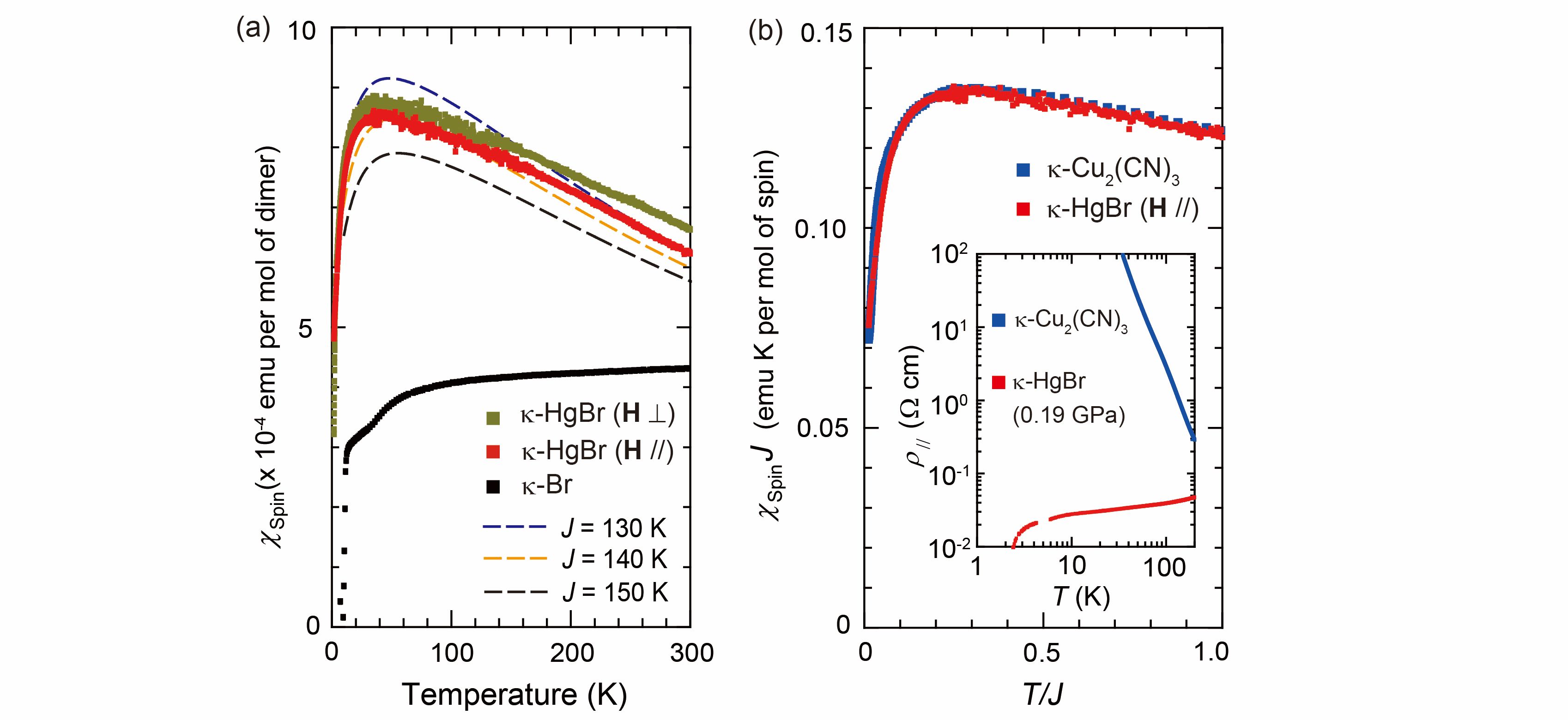}
\caption{\label{fig:wide}
(a) Temperature dependences of the spin susceptibility $\chi_{\rm spin}$ of the doped compound $\kappa$-(BEDT-TTF)$_4$Hg$_{2.89}$Br$_8$ and a half-filled compound $\kappa$-(BEDT-TTF)$_2$Cu[N(CN)$_2$]Br (denoted as $\kappa$-HgBr and $\kappa$-Br, respectively). $\chi_{\rm spin}$ of $\kappa$-(BEDT-TTF)$_4$Hg$_{2.89}$Br$_8$ under magnetic fields parallel and perpendicular to the conducting layers is indicated by red and green points, respectively. The broken lines represent the numerical curves obtained by the series expansion of the triangular-lattice Heisenberg model with a $J$ value of 130–150 K. Reprented from Ref. \cite{oike2017anomalous} under the creative commons license. (b) Scaled $\chi_{\rm spin}$ of $\kappa$-(BEDT-TTF)$_4$Hg$_{2.89}$Br$_8$ and $\kappa$-(BEDT-TTF)$_2$Cu$_2$(CN)$_3$ (denoted as $\kappa$-Cu$_2$(CN)$_3$). $J\chi_{\rm spin}$ is plotted against $T/J$, where $\chi_{\rm spin}$ is defined per mole of spin and $J$ is in units of $k_{\rm B}$. The inset shows the in-plane resistivity of the two compounds. Reprented from Ref. \cite{oike2017anomalous} under the creative commons license.}
\end{figure*}
%%%%%%%%%%%%%%%%%%%Fig12_wide

Bandwidth-controlled metal-insulator transitions observed in non-doped organic compounds and band-filling-controlled metal-insulator transitions widely studied in inorganic materials like cuprates have been discussed in parallel \cite{mckenzie1997similarities}. Nevertheless, we note that they also have different features; e. g., the metal-insulator transitions in the former is of first order \cite{kagawa2005unconventional,kanoda2011mott}, whereas in the latter, the transition appears continuous in many cases \cite{imada1998metal}. $\kappa$-Hg$_{2.89}$Br$_8$ offers an unique platform to elucidate the interplay between Mottness and itineracy behind the two kinds of metal-insulator transitions. Furthermore, the metallic states that are qualitatively different from a conventional Fermi liquid due to non-perturbative effects of electronic correlations also appear in heavy Fermion materials, and how to conprehensively understand such strange metallicity is an intriguing issue \cite{cai2020dynamical, paschen2021quantum, nguyen2021superconductivity, yang2021doping} as discussed in Chapter 5.

\section{Spin liquidity in the doped organic conductor}
In a conventional Fermi liquid state, conduction electrons exhibit Pauli-paramagnetism, obeying Fermi statistics whereas electrons in a Mott insulator lose the charge degrees of freedom and behave like a localized spin system. Thus, Mottness is closely related to magnetism. Then, what does magnetism of doped Mott insulators look like? A doped Mott insulator is similar to a Mott insulator in terms of Mottness, whereas it is similar to a conventional Fermi liquid in terms of metallicity. A well-known manifestation possessing contradictory properties is a spin gap in the metallic state as observed in high-$T_{\rm c}$ cuprates. What makes $\kappa$-Hg$_{2.89}$Br$_8$ even more unique is that the parent Mott insulator can be a QSL instead of antiferromagnets for the cuprates. This chapter describes the magnetism of the triangular-lattice doped Mott insulator $\kappa$-Hg$_{2.89}$Br$_8$. 

\subsection{Spin-charge separation}
Fig. 12 compares the spin susceptibility $\chi_{spin}$ of three $\kappa$-type BEDT-TTF compounds, the half-filled metal $\kappa$-(BEDT-TTF)$_2$Cu[N(CN)$_2$]Br, the Mott-insulating QSL candidate $\kappa$-(BEDT-TTF)$_2$Cu$_2$(CN)$_3$, and the present system $\kappa$-Hg$_{2.89}$Br$_8$ \cite{oike2017anomalous,shimizu2003spin,skripov1989anomalous,kanoda2006metal}. Surprisingly, the $\chi_{spin}$ behaviors of $\kappa$-Hg$_{2.89}$Br$_8$ and $\kappa$-(BEDT-TTF)$_2$Cu$_2$(CN)$_3$ makes no difference when $\chi_{spin}$ and $T$ are scaled to an exchange interaction, $J$ (Fig. 12(b)) while they behave contrastingly in conductivity (Inset of Fig. 12(b)). The $\chi_{spin}$ of $\kappa$-(BEDT-TTF)$_2$Cu$_2$(CN)$_3$ is well reproduced by a series expansion of the triangular-lattice Heisenberg model with a $J$ of 250 K \cite{elstner1993finite}. What is peculiar is that  the $\chi_{spin}$ of $\kappa$-Hg$_{2.89}$Br$_8$ is also well reproduced by the triangular-lattice Heisenberg model down to $T \sim 0.2 J$ or lower with a $J$ of 140 K as if it hosts localized spins in spite of the metallicity. Note that the isostructural metallic compounds with half-filled bands, $\kappa$-(BEDT-TTF)$_2$Cu[N(CN)$_2$]Br and $\kappa$-(BEDT-TTF)$_2$Cu(NCS)$_2$, exhibit a Pauli paramagnetic spin behavior. 

We comment on the scaling in terms of $J$. In Fig. 12(b), the horizontal axis is the temperature divided by $J$, meaning that thermal energy is measured in units of $J$, and the vertical axis is the spin susceptibility multiplied by $J$, meaning that the Zeeman energy is measured in units of $J$, namely, the magnetic field is measured in units of $\mu/J$, where $\mu$ is the magnetic moment of an electron spin. In addition, the $\chi_{spin}J$ value is displayed in units of emu K per mole of spin to see how a spin is polarized under the scaled magnetic field. Because 11\% carrier doping reduces the number of spins from 1 to 0.89 per a BEDT-TTF dimer site, the value per mole of dimer is multiplied by 0.89 for $\kappa$-Hg$_{2.89}$Br$_8$ and 1 for $\kappa$-(BEDT-TTF)$_2$Cu$_2$(CN)$_3$. The magnetic similarity and conductive dissimilarity between two systems are a clear indication of spin–charge separation in that carrier doping does not affect the spin sector. 

A prominent feature in the scaled susceptibility is a sharp decrease at low temperatures below $\sim$ 0.1 $J$, which is out of the relevant temperature range of the series expansion. Although a triangular-lattice Heisenberg model results in 120-degree long-range order at the lowest temperature, magnetic ordering is absent in these compounds \cite{satoh2009musr}. Then, the sharp decrease in the susceptibility is not a simple indication of the growth of the 120-degree spin correlation, but may signify the growth of quantum mechamical degeneracy inherent in a QSL. 

\subsection{Spin excitations and interlayer charge transport}
In the possible doped QSL state of $\kappa$-Hg$_{2.89}$Br$_8$, the charge and spin degrees of freedom seem to be separately excited in a conducting layer as discussed in the preceeding section. In this situation, let's consider the interlayer transfer of electrons. It should accompany the simultaneous transfer of charge and spin because they are usually inseparable as the constituent degrees of freedom of an electron. This means that the interlayer charge transport requires its recoupling with the separated spin degrees of freedom. Thus, even in a metallic state within the layers, it is nontrivial whether the our-of-plane resistivity $\rho_{\perp}$ is metallic.

%%%%%%%%%%%%%%%%%%%Fig13
\begin{figure}
\includegraphics[width=86mm]{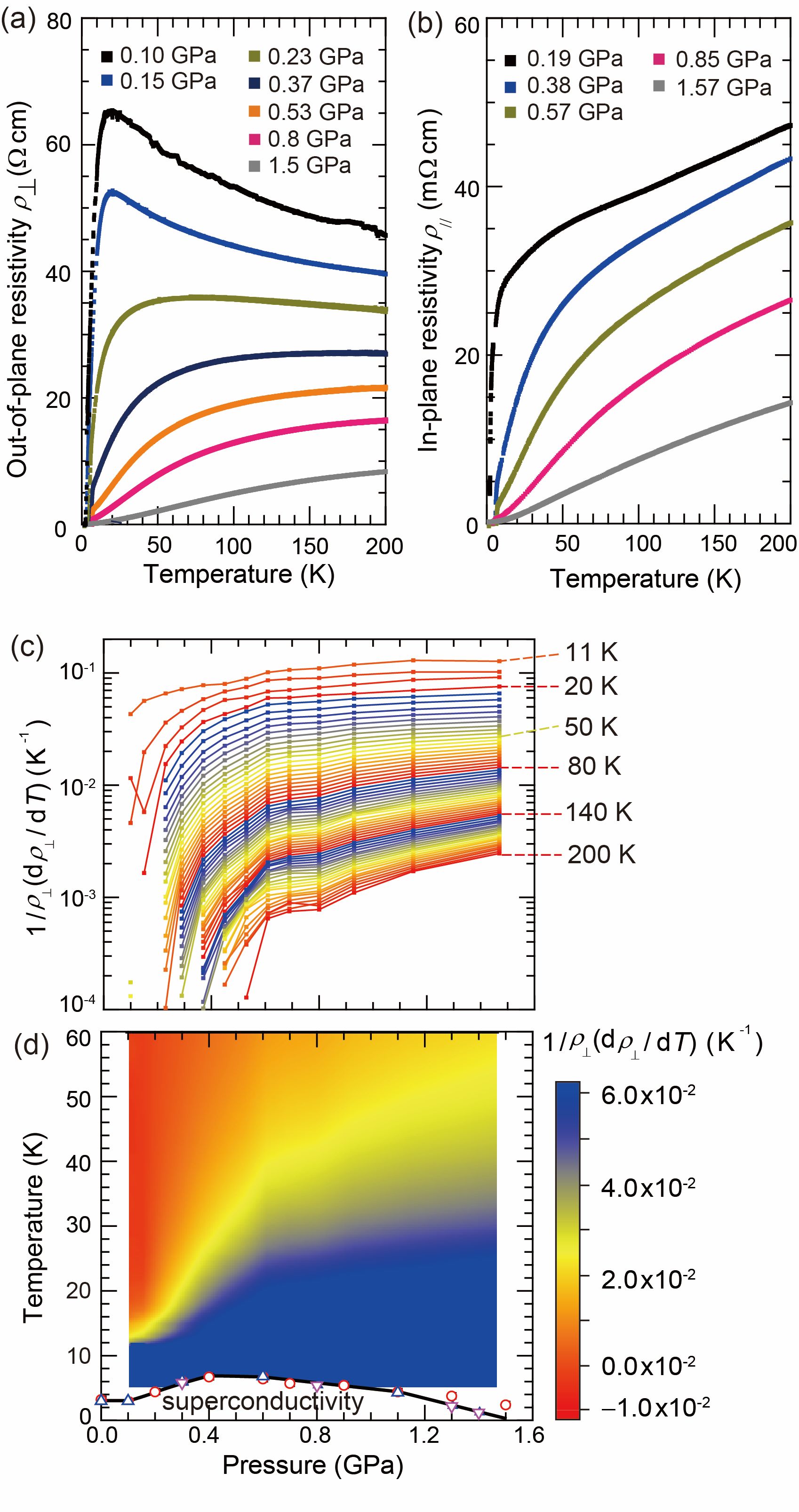}
\caption{\label{fig}
(a, b) Temperature dependences of the out-of-plane resistivity $\rho_{\perp}$ (a) and in-plane resistivity $\rho_{//}$ (b) at several pressures. Reprented from Ref. \cite{oike2017anomalous} under the creative commons license. (c) Pressure dependence of the normalized temperature derivative of the out-of-plane resistivity 1/$\rho_{\perp}$($d$$\rho_{\perp}$/$dT$). Reprented from Ref. \cite{oike2017anomalous} under the creative commons license. (d) Contour plot of 1/$\rho_{\perp}$($d$$\rho_{\perp}$/$dT$) in the pressure–temperature plane. Reprented from Ref. \cite{oike2017anomalous} under the creative commons license.}
\end{figure}
%%%%%%%%%%%%%%%%%%%Fig13

%%%%%%%%%%%%%%%%%%%Fig14
\begin{figure}
\includegraphics[width=86mm]{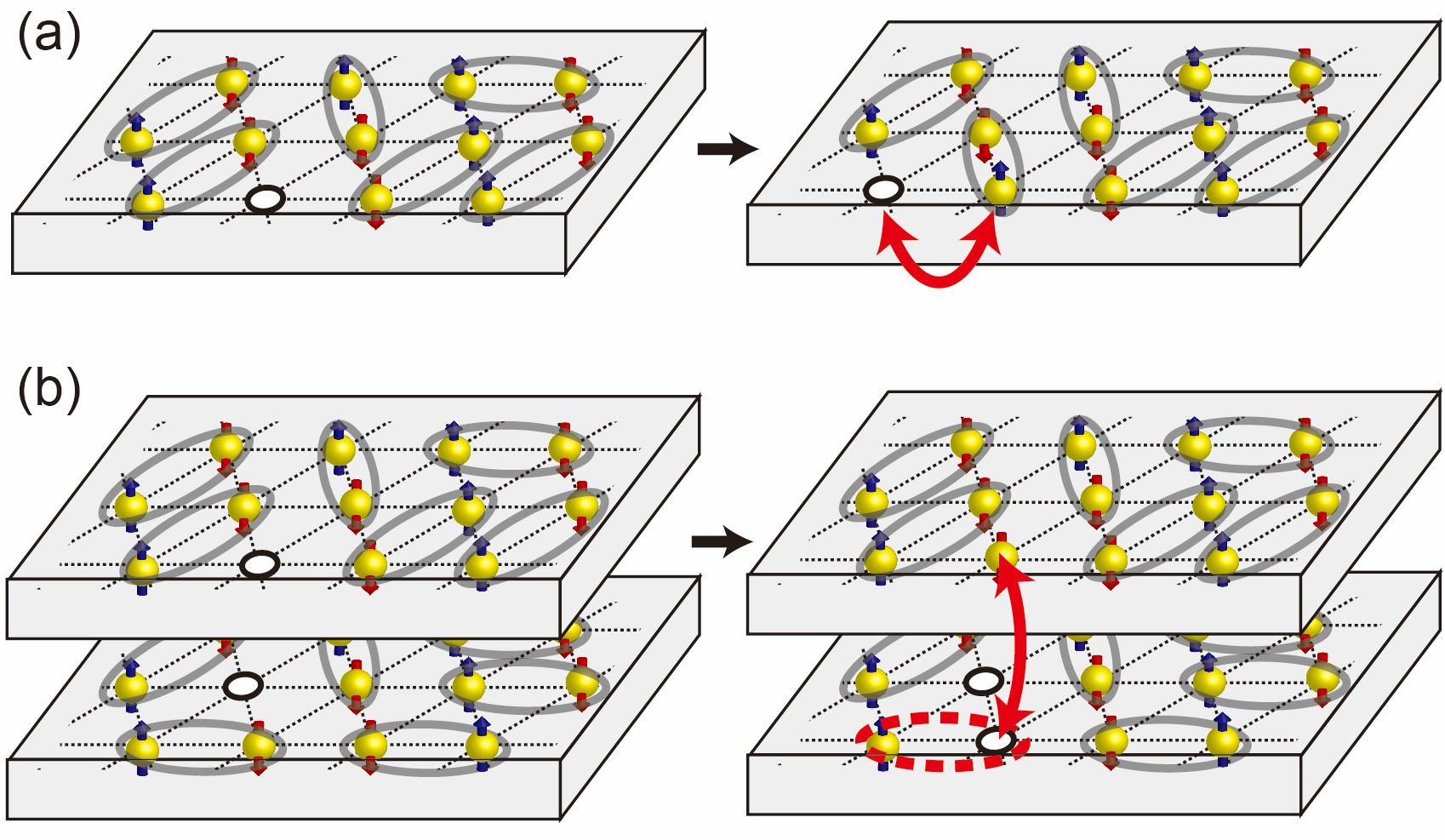}
\caption{\label{fig}
(a, b) Schematic figure of in-plane charge transport (a) and out-of-plane charge transport (b). When spin singlet states resonate in a plane, in-plane charge transport does not involve the destruction of a spin singlet, resulting in metallic conduction. However, out-of-plane charge transport requires the destruction of a spin singlet, and thus $\rho_{\perp}$ can exhibit nonmetallic behavior.}
\end{figure}
%%%%%%%%%%%%%%%%%%%Fig14

Figs. 13(a) and 13(b) compare the $\rho_{\perp}$ and $\rho_{//}$ of $\kappa$-Hg$_{2.89}$Br$_8$. Remarkably, $\rho_{\perp}$ exhibits nonmetallic behavior over a wide temperature range and sharply turns to metallic below 20 K at 0.1 and 0.15 GPa \cite{oike2017anomalous}, while $\rho_{//}$ is metallic in the whole pressure-temperature range investigated. The nonmetallic behavior of $\rho_{\perp}$ disappears at pressures, and above 0.8 GPa, $\rho_{\perp}$ obeys the temperature-square law of a Fermi liquid as $\rho_{//}$ does \cite{oike2017anomalous}. The contour map of the normalized temperature derivative of $\rho_{\perp}$, 1/$\rho_{\perp}$(d$\rho_{\perp}$/d$T$), captures the overall behavior of $\rho_{\perp}$ in the pressure-temperature diagram (Fig. 13(d)), highlighting three distinctive pressure regions; that is, $P <$ 0.3 GPa, 0.3 $< P <$ 0.6 GPa and $P >$ 0.6 GPa. The high-pressure region corresponds to the FL region identified by the temperature-square dependences of both $\rho_{//}$ and $\rho_{\perp}$. In the low-pressure region, the nonmetal-to-metal crossover of $\rho_{\perp}$ sharply occurs at low temperatures with the temperature-linear behavior of $\rho_{//}$. In the intermediate-pressure region, the crossover temperature increases towards the high-pressure region and the nonmetallic behavior is less pronounced.

Of particular interest is the contrasting behavior of $\rho_{//}$ and $\rho_{\perp}$ in the low-pressure region. In the non-doped compound $\kappa$-(BEDT-TTF)$_2$Cu[N(CN)$_2$]Cl, $\rho_{//}$ and $\rho_{\perp}$ simultaneously change from a metallic to an insulating temperature dependence across the critical pressure of the metal-insulator transitions \cite{ito1996metal}. Thus, the contrasting behavior is characteristic of $\kappa$-Hg$_{2.89}$Br$_8$. As its possible origin, in-plane spin correlations can differently affect $\rho_{//}$ and $\rho_{\perp}$ as follows. For example, if spin and charge degrees of freedom are separated, the interlayer electron transfer necessitates their simultaneous hopping, which should be a higher-order event. Or if spins form some entangled states like the RVB state \cite{anderson1973resonating,anderson1987resonating,savary2016quantum}, the out-of-plane charge transport requires the breaking of the resonating bonds with an excitation gap while the in-plane one does not (Fig. 14). It is quite interesting how the energy distribution of spin excitations in the presumable QSL relates to that of charge excitations as theoretically argued as dimensional decoupling at continuous quantum critical Mott transitions \cite{zou2016dimensional}. 

This argument on the nonmetallic behavior of $\rho_{\perp}$ provokes further interest in why it changes to metallic behavior under pressure or upon cooling. The pressure induces the bandwidth-controlled Mottness transition from a strange metal to a Fermi liquid as discussed in Chapter 4. In a Fermi liquid, elementary excitations are carried by quasi-particles with charge and spin combined, whose transport is essentially the same in the intra and inter-layer directions. Interestingly, the nonmetallic behavior crosses over to the metallic one upon cooling in the low-pressure region. This implies that spin and charge are coupled in a different manner from a conventional Fermi liquid state because the temperature-linear behavior of $\rho_{//}$ persists down to the lowest temperature. Thus, it would be an intriguing issue to examine how the separated spin and charge degrees of freedom recouple to form unknown elementary excitations at low temperatures.

\subsection{Exotic excitations in the doped quantum spin liquid}
The specific heat measurements underscore the unique excitations in $\kappa$-Hg$_{2.89}$Br$_8$ \cite{naito2003low, naito2005anomalous,yamashita2005drastic}. Although the one-dimensional phonons due to the incommensurate Hg sublattice complicate the analysis that separates the electronic contribution from the phonon contribution and the Schottky anomaly, the electronic specific heat coefficient $\gamma$ is evaluated to be 55 mJ mol$^{-1}$ K$^{-2}$ \cite{naito2005anomalous}. How to address this value consistently with the non-Fermi liquid behaviors of $\kappa$-Hg$_{2.89}$Br$_8$ is an open issue of interest. We note that this value is extraordinarily large among organic conductors; e.g., $\gamma$ = 25 and 22 mJ mol$^{-1}$ K$^{-2}$ for the half-filled metals, $\kappa$-(BEDT-TTF)$_2$Cu(NCS)$_2$ and $\kappa$-(BEDT-TTF)$_2$Cu[N(CN)$_2$]Br, respectively \cite{naito2005anomalous}. This suggests that an unusually large number of gapless excitations are degenerate in $\kappa$-Hg$_{2.89}$Br$_8$. The Wilson ratio, ($\pi^2$/3)($\chi_{spin}$/$\mu_{B}^2$)/($\gamma$/$k_B^2$), gives an additional insight to the nature of low-temperature thermal excitations. $\kappa$-Hg$_{2.89}$Br$_8$ exhibits an anomalous value of 0.6. This value is extraordinartily smaller than the values for the half-filled systems, 0.95 and 0.98, respectively, which are typical values for Fermi liquids \cite{wilson1975renormalization,yamada1975perturbation}. The low Wilson ratio implies that gapless excitations detected in the specific heat measurements do not fully contribute to spin polarization unlike in Fermi liquids. We compare the $\gamma$ value and Wilson ratio of $\kappa$-Hg$_{2.89}$Br$_8$ with those of its undoped version $\kappa$-(BEDT-TTF)$_2$Cu$_2$(CN)$_3$. The $\gamma$ value of 13 mJ mol$^{-1}$ K$^{-2}$ for $\kappa$-(BEDT-TTF)$_2$Cu$_2$(CN)$_3$ is discussed as a hallmark of the spinon Fermi surfaces in conjunction with the acceptable Wilson ratio, 1.6, which is comparable to another QSL candidate, Me$_3$EtSb[Pd(dmit)$_2$] \cite{itou2008quantum,yamashita2008thermodynamic,yamashita2011gapless}. Assuming the spin-charge separation, one may independently consider the spin and charge contributions to the $\gamma$ value of $\kappa$-Hg$_{2.89}$Br$_8$.
Given that the spin sector has the common Wilson ratio, 1.6, for these QSL candidates as a spinon property, the spin contribution to the $\gamma$ value is estimated by dividing the $\chi_{spin}$ value of 5 $\times$ 10$^{-2}$ emu mole$^{-1}$ by 1.6 and yields 22 mJ mol$^{-1}$ K$^{-2}$, which explains only a part of the observed $\gamma$ value. Then, the remaining large value, 33 mJ mol$^{-1}$ K$^{-2}$, can be attributed to the charge sector formed by 11\% doped holes although it is questionable whether the charge sector is fermionic. Anyway, the spin-charge separation, in which the charge sector does not contribute to $\chi_{spin}$, offers one possible explanation to the low Wilson ratio in $\kappa$-Hg$_{2.89}$Br$_8$. 

To view the present results with a more broad perspective, we compare the present single-orbital material $\kappa$-Hg$_{2.89}$Br$_8$ with multi-orbital metallic QSL candidates Pr$_2$Ir$_2$O$_7$ and CePdAl, in which localized and itinerant electrons coexist \cite{senthil2003fractionalized,burdin2002heavy,si2006global,coleman2010frustration,kim2013spin,pixley2014quantum,tokiwa2014quantum, zhao2019quantum}. In these materials, spatially localized orbitals, e. g., the $f$ orbital of a rare earth element, hybridize with spatially extended orbitals. When the hybridization is strong, a Fermi liquid state with a heavy electron mass appears. In a weakly hybridized regime, however, a band-selective Mott transition called Kondo breakdown is proposed to result in a certain class of materials \cite{si2010heavy, prochaska2020singular}, where only the $f$ electrons fall into the Mott insulating state. Kondo breakdown is similar to the Mottness transition in $\kappa$-Hg$_{2.89}$Br$_8$ in that two kinds of metallic states are distinguished by Mottness, and is also relevant to multi-orbital organic conductors \cite{takagi2017single, briere2018interplay, takagi2020multiorbital, kobayashi2021single}.

When the localized $f$-electron spins are subject to spin frustration, e. g., in the Shastry–Sutherland lattice and pyrochlore lattices, a metallic state with a QSL nature can emerge as theoretically suggested \cite{pixley2014quantum,tokiwa2014quantum, vojta2018frustration, wang2022z}. In these metallic spin liquid states of multiorbital systems, spin and charge sectors are disentangled at high temperature and weakly hybridized to form an anomalous metallic state, which is mentioned as a fractionalized Fermi liquid, at low temperatures \cite{senthil2003fractionalized}. The similar temperature dependence of entanglement between spin and charge sectors may explain the appearance of metallic behavior of $\rho_{\perp}$ of $\kappa$-Hg$_{2.89}$Br$_8$ in the low-pressure region as discussed in the previous section. In addition, $\kappa$-Hg$_{2.89}$Br$_8$ exhibits non-Fermi liquid behavior in a finite pressure range, which is similar to the metallic QSL candidate in the multiorbital system CePdAl \cite{zhao2019quantum, ramires2019frustration}. This similarity implies that, in the presence of QSL, electronic fluids form non-Fermi liquid phases common to single orbital and multiorbital systems. Despite the similarities with multiorbital systems, however, it should be noted again that $\kappa$-Hg$_{2.89}$Br$_8$ is a single band system; Only one of the separated spin and charge degrees of freedom here can be fermionic while those in the above-mentioned multi-orbital materials are both fermions. Moreover, the effect of carrier doping on frustrated spin systems has recently been investigated in cold-atom systems \cite{xu2022doping}. Further investigations on similarities and individualities among those systems are expected to offer an extended view on the relation between the QSL nature and non-Fermi liquid nature.

\section{Superconductivity in the doped organic conductor}
Superconductivity often appears when doping or pressurizing Mott insulators, as represented by high-$T_{\rm c}$ cuprates, iron-based compounds, fullerides and $\kappa$-type BEDT-TTF compounds, implying an intimate relationship between superconductivity and Mottness. However, because the energy scale of on-site Coulomb repulsion is much larger than that of cooper pairing, we need concepts that connect the phenomena of this energetic hierarchy. Mottness can be varied while metallicity is maintained in pressure experiments on $\kappa$-Hg$_{2.89}$Br$_8$ as mentioned in Chapter 4, and thus $\kappa$-Hg$_{2.89}$Br$_8$ is a suitable material for investigating the role of Mottness in the emergence of superconductivity. Furthermore, $\kappa$-Hg$_{2.89}$Br$_8$ also enables to clarify the relationship between spin liquidity and superconductivity, which has been of theoretical interest \cite{anderson1987resonating, jiang2021superconductivity, lopez2022topological}, because the QSL nature is relevant at low pressures as mentioned in Chapter 5. This chapter describes the pressure variation in the superconducting properties of the doped Mott insulator $\kappa$-Hg$_{2.89}$Br$_8$.

\subsection{Pressure dependence of the superconducting transition temperature}
Superconductivity in $\kappa$-Hg$_{2.89}$Br$_8$ was first reported in 1987 by Lyubovskaya et al., \cite{lyubovskaya1987organic} and the nonmonotonous pressure dependence of the superconducting transition temperature $T_{\rm c}$ was reported in 1992 by Bud’ko et al. \cite{bud1992anomalous}, who also found a nonmetallic phase emerging at high pressures; however, it has not been reproduced. Contrastingly, there was observed a resistivity drop reminiscent of superconductivity at high pressures exceeding 3 GPa \cite{taniguchi2007anomalous}; however, this also has not been reproduced. Presumably, these observations are due to poor hydrostaticity of pressure or strain in the sample. The ac susceptibility measurements made afterward confirmed superconducting transitions in a pressure range of 0-1.7 GPa with a dome-shaped pressure dependence of $T_{\rm c}$ (Fig. 15) \cite{oike2015pressure}. The superconducting diamagnetism at ambient pressure was smaller than those at pressures above 0.3 GPa, indicating that the superconducting volume, at least, at ambient pressure is fractional; indeed, the specific heat measurements suggest the superconducting volume fraction of 30-50 \% at ambient pressure \cite{naito2005anomalous}. The diamagnetic response at 1.4 K increases with pressure, reaching twice as much at 0.3 GPa, and saturates in pressure and temperature variations, suggesting that the entire sample is superconducting above 0.3 GPa consistently with the recent thermodynamic measurements under pressure \cite{matsumura2022thermodynamic}. (Very recently, a few crystals with a full superconducting volume fraction at ambient pressure were successfully synthesized \cite{sari2023superconductivity, wakamatsu2022reduced}.)

%%%%%%%%%%%%%%%%%%%Fig15
\begin{figure}
\includegraphics[width=86mm]{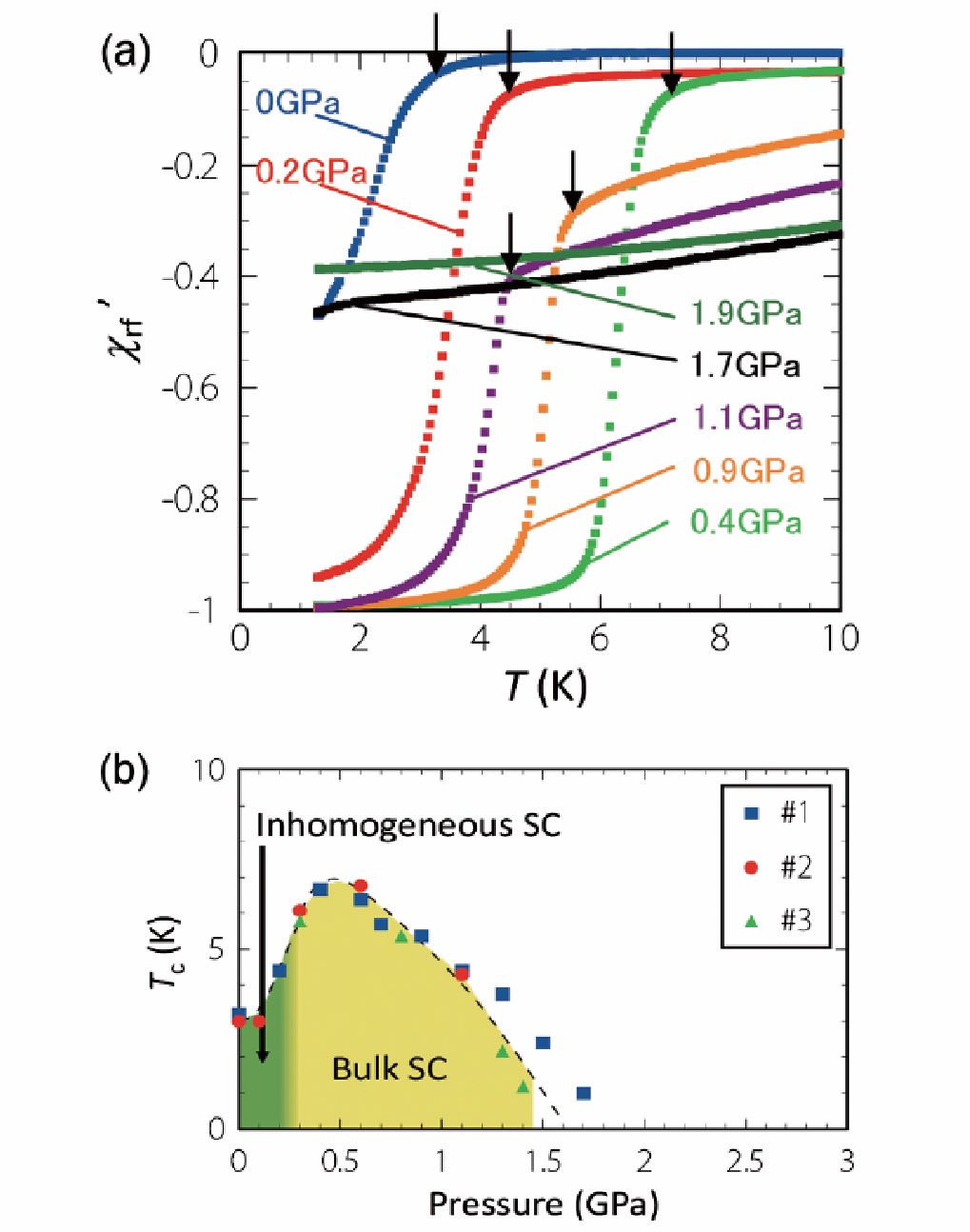}
\caption{\label{fig}
(a) Temperature dependence of ac susceptibility in a MHz frequency range $\chi'_{\rm rf}$ of $\kappa$-(BEDT-TTF)$_4$Hg$_{2.89}$Br$_8$ under pressure. The arrows indicate the superconducting (SC) transition temperature $T_c$. The finite value of $\chi'_{\rm rf}$ above $T_c$ is due to the skin effect, which enables contactless conductivity measurements, as described in Chapter 4. Reprinted from Ref. \cite{oike2015pressure}. \copyright{2015 the American Physical Society}. (b) Pressure dependence of $T_c$ probed by ac susceptibility measurements. At ambient pressure and 0.2 GPa, the SC diamagnetism is not perfect, indicating that a part of the sample does not exhibit an SC transition consistent with the specific heat measurement at ambient pressure. Although the origin of inhomogeneity in a sample is not clarified, the pressure application results in a bulk superconductivity above 0.3 GPa. Reprinted from Ref. \cite{oike2015pressure}. \copyright{2015 the American Physical Society}.}
\end{figure}
%%%%%%%%%%%%%%%%%%%Fig15

%%%%%%%%%%%%%%%%%%%Fig16_wide
\begin{figure*}
\includegraphics[width=172mm]{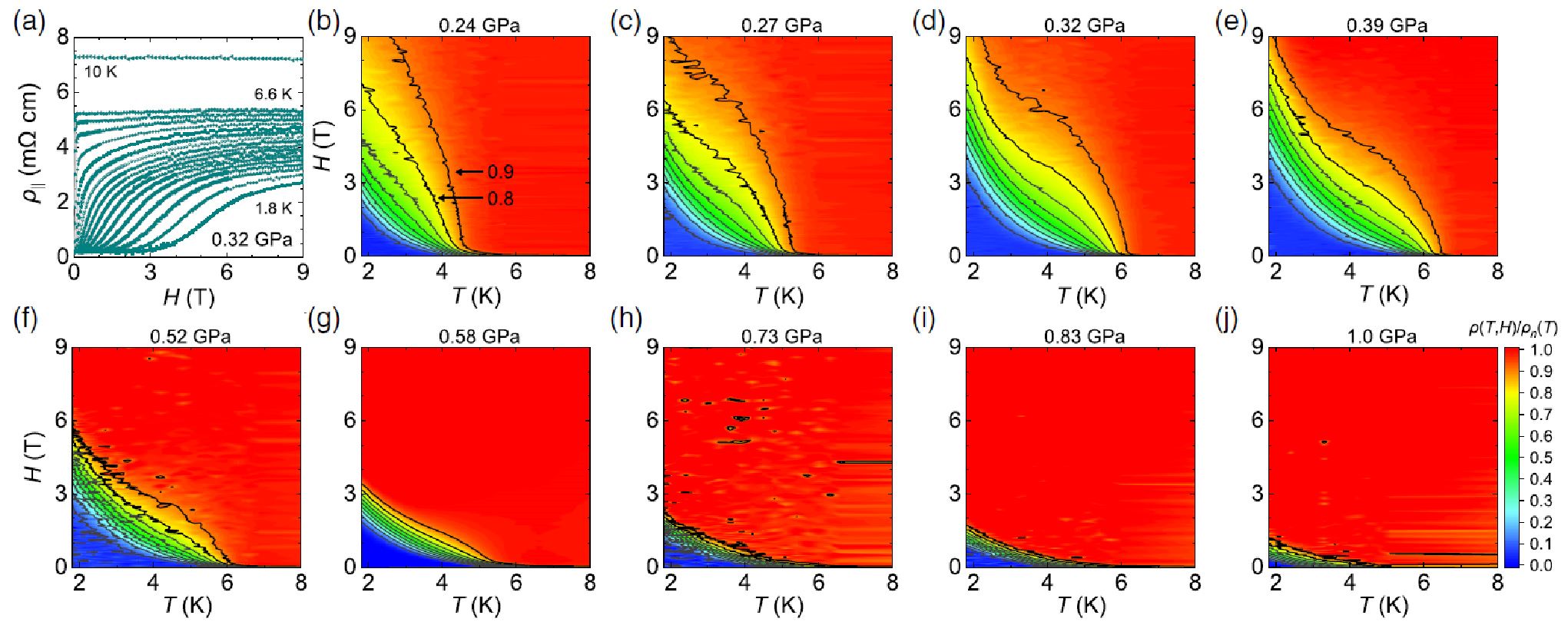}
\caption{\label{fig:wide}
(a) Magnetic field $H$ dependence of in-plane resistivity $\rho_{//}$ at a pressure of 0.32 GPa, where the field is applied perpendicular to the conducting layers. Reprented from Ref. \cite{suzuki2022mott} under the creative commons license. (b)–(j) Contour plots of the normalized in-plane resistivity, $\rho_{//}(T, H)/\rho_{\rm n}$, in the temperature-field ($T$-$H$) plane, where $\rho_{\rm n}$ is the normal state resistivity with the form of $\rho_{o} + AT^{\alpha}$. The values of $\rho_{o}$, $A$ and $\alpha$ are obtained by fitting the functional form to the high-field data. The solid lines indicate the contours of $\rho_{//}(T, H)/\rho_{\rm n}$ = 0.1, 0.2, 0.3, 0.4, 0.5, 0.6, 0.7, 0.8, and 0.9. The red and blue areas indicate the normal and superconducting phases, respectively. Reprented from Ref. \cite{suzuki2022mott} under the creative commons license.}
\end{figure*}
%%%%%%%%%%%%%%%%%%%Fig16_wide

The pressure dependence of $T_{\rm c}$ shows a peak at a pressure of approximately 0.5 GPa, which corresponds to the pressure range of the bandwidth-controlled Mottness transition, as described in Chapter 4. In non-doped $\kappa$-type BEDT-TTF compounds \cite{kanoda1997electron}, $T_{\rm c}$ decreases with increasing pressure in a Fermi liquid phase, commonly to $\kappa$-Hg$_{2.89}$Br$_8$ above 0.5 GPa. These behaviors indicate that the energy scale of Cooper pairing in a Fermi liquid phase increases toward a Mottness transition. When the Mottness is further increased and $\kappa$-Hg$_{2.89}$Br$_8$ enters into the low-pressure non-Fermi liquid regime, $T_{\rm c}$ turns into a decrease. Contrastingly, the pairing interaction appears to increase with decreasing pressures as will be discussed in the section 6-C. This puzzling behavior is likely related to the interlayer decoupling, which works against the three-dimensional ordering, in the non-Fermi liquid regime as we discussed in the section 5-B.

\subsection{Mottness-Driven BEC-BCS Crossover}
As seen in the chapter 4, the nature of carriers responsible for electrical conduction on the low-pressure side of the $T_{\rm c}$ dome differs from that on the high-pressure side. Thus, one may expect something different in Cooper paring between the two pressure regimes. On the high-pressure side, the quasiparticles of the Fermi liquid form Cooper pairs; however, in a state where conventional quasi-particles are not well defined, the formation of Cooper pairs is difficult to be described. A critical magnetic field probes the spatial extent of the Cooper pair and thus its pressure variation gives insight into this issue.

%%%%%%%%%%%%%%%%%%%Fig17
\begin{figure}
\includegraphics[width=86mm]{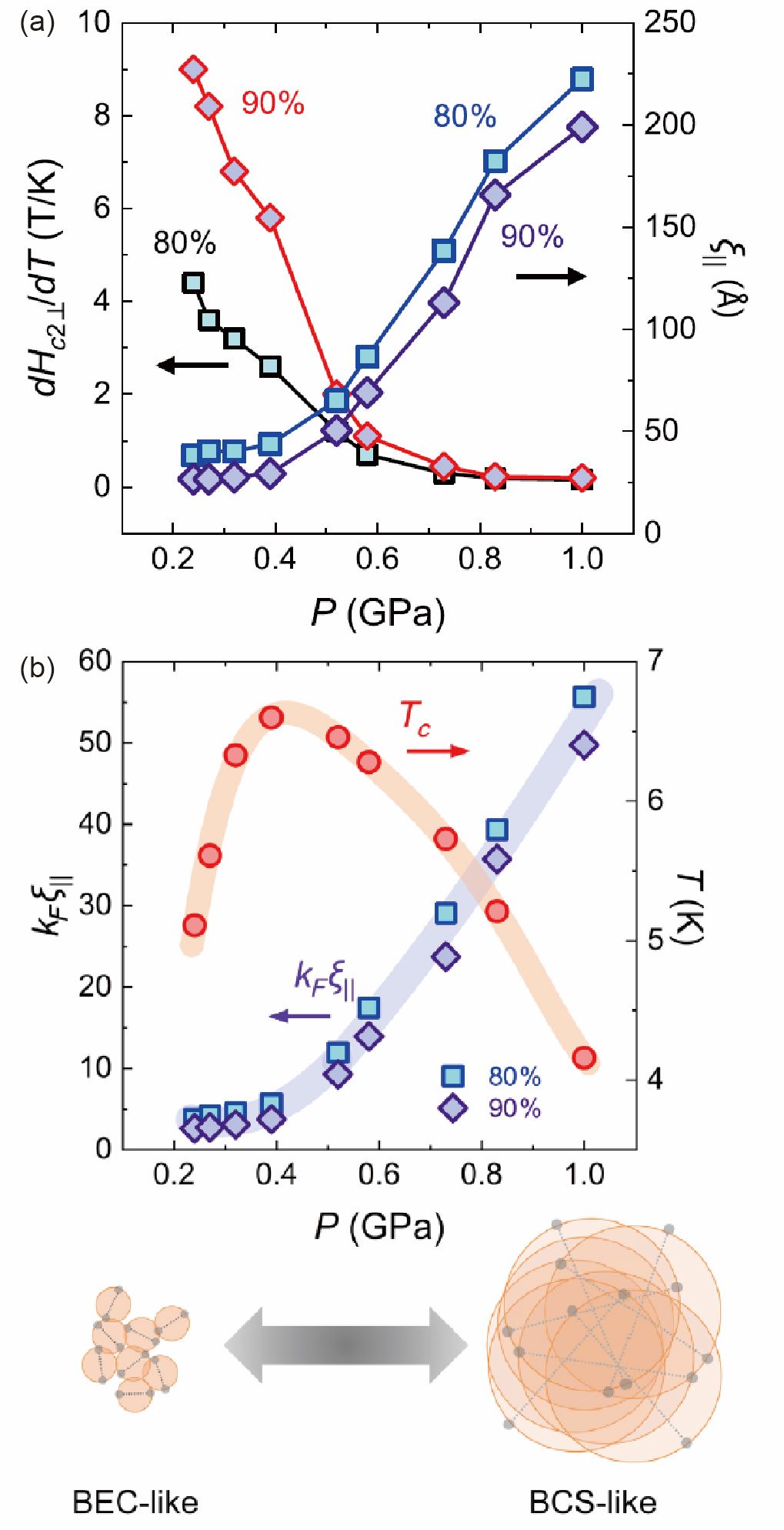}
\caption{\label{fig}
(a) Pressure dependences of d$H_{c2\perp}$/d$T_{\rm c}$ and $\xi_{//}$. (b) Pressure dependences of $k_{\rm F}$$\xi_{//}$ and superconducting (SC) transition temperature $T_c$, where $k_{\rm F}$ and $\xi_{//}$ are the Fermi wavenumber and a SC coherence length, respectively. The value of $k_{\rm F}$$\xi_{//}$ monotonically increases from 3 to 50 as a function of pressure. As schematically depicted, a small value of $k_{\rm F}$$\xi_{//}$ at low pressures indicates that a few Cooper pairs overlap similarly to a Bose-Einstein Condensation (BEC)-like superconductivity, and a large value of $k_{\rm F}$$\xi_{//}$ at high pressures indicates that many Cooper pairs overlap as in a Bardeen-Cooper-Schrieffer (BCS)-like superconductivity. Reprented from Ref. \cite{suzuki2022mott} under the creative commons license.}
\end{figure}
%%%%%%%%%%%%%%%%%%%Fig17

%%%%%%%%%%%%%%%%%%%Fig18_wide
\begin{figure*}
\includegraphics[width=172mm]{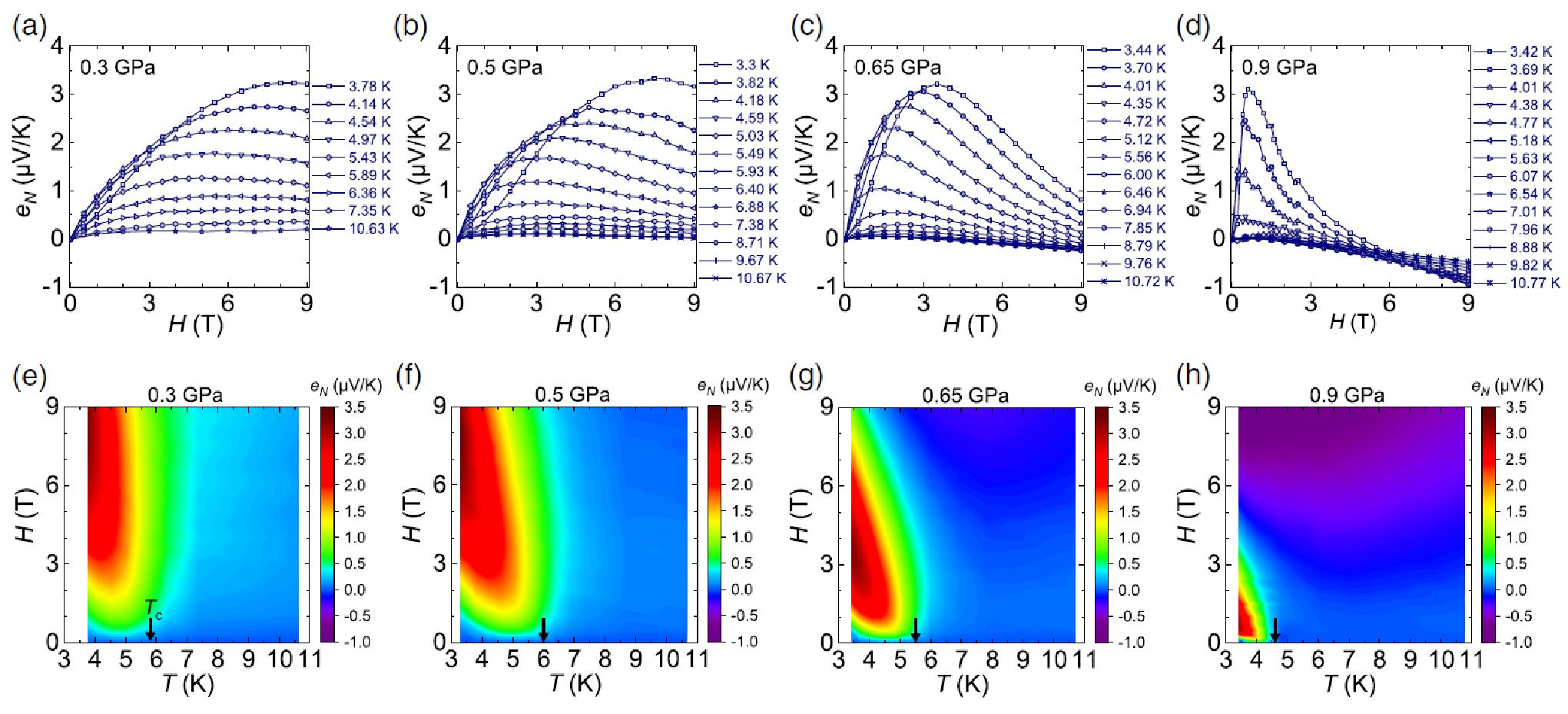}
\caption{\label{fig:wide}
(a)–(d) Magnetic field ($H$) dependence of the Nernst signal $e_{\rm N}$ at several fixed temperatures under pressure, where the field is applied perpendicular to the conducting layers. (e)–(h) Contour plots of $e_{\rm N}$ in the temperature–field ($T$,-$H$) plane. The arrows indicate superconducting transition temperatures at $H=0$. Reprented from Ref. \cite{suzuki2022mott} under the creative commons license.}
\end{figure*}
%%%%%%%%%%%%%%%%%%%Fig18_wide

Fig. 16 shows contour plots of the normalized in-plane resistivity $\rho_{//}$ under perpendicular magnetic fields at several pressures \cite{suzuki2022mott}. Although superconductivity is broken at relatively low fields of a few Tesla in a high-pressure range above 0.5 GPa, superconductivity persists to high magnetic fields in a low-pressure range below 0.5 GPa. The contrasting magnetic field dependencies indicate that the upper critical field $H_{c2\perp}$ does not scale to $T_{\rm c}$, consistently with the unusually large critical fields at ambient pressure \cite{ohmichi1999superconducting, shimojo2003upper, imajo2021extraordinary}. $H_{c2\perp}$ is approximately related with the in-plane coherence length $\xi_{//}$ through $H_{c2\perp}$ = $\phi_o$/2$\pi\xi_{//}^2$ \cite{tinkham2004introduction}, where $\phi_o$ is the flux quantum. However, the determination of $H_{c2\perp}$ from magnetoresistance require a particular caution for highly two-dimensional superconductors. When a magnetic field is applied perpendicular to layered superconductors, finite resistivity is often caused by vortex flow in the vortex liquid state even below $H_{c2\perp}$ \cite{tinkham2004introduction}, as observed in this system. In such a case, $H_{c2\perp}$ is better characterized by an initial drop in resistivity, e.g. 80\% or 90\% transition points [$\rho_{//}$($T$, $H$)/ $\rho_{n}$($T$)  = 0.8 and 0.9], which are indicated by the bold contours in Figs. 16(b)–16(j). Then, the GL coherence length $\xi_{//}$ is evaluated by extrapolating the temperature slope of $H_{c2\perp}$ near $T_{\rm c}$ to absolute zero. The pressure dependence of $\xi_{//}$ is shown in Fig. 17(a). $\xi_{//}$ at 0.4 GPa is as short as 3 nm, which is the length of a few unit cells, and increases with pressure to a value of ~20 nm at 1 GPa, which is a close value to the non-doped $\kappa$-type BEDT-TTF compounds \cite{suzuki2022mott}.

Having clarified the pressure variation of $\xi_{//}$, it is of interest to know how many Cooper pairs spatially overlap with one another. Whereas more than 10,000 pairs overlap in typical Bardeen-Cooper-Schrieffer (BCS) superconductors, such as Sn or Pb, the overlapping of only a few pairs or less takes on the nature of a Bose-Einstein Condensation (BEC) \cite{uemura1991basic,uemura1991magnetic,uemura2003superfluid,hartke2022direct}. In the first approximation, $\xi_{//}$ is regarded as the size of the Cooper pair. A carrier density is associated with a length scale 1/$k_{\rm F}$, the inverse of the Fermi wavenumber. Thus, $k_{\rm F}$$\xi_{//}$ measures the degree of the pair overlapping and serves as an indicator of the BEC/BCS feature \cite{pistolesi1994evolution}. The $k_{\rm F}$ values of $\kappa$-Hg$_{2.89}$Br$_8$ under pressure are estimated from the Hall coefficient by assuming a cylindrical Fermi surface, and the $k_{\rm F}$$\xi_{//}$ values are obtained as shown in Fig. 17(b) \cite{suzuki2022mott}. At high pressures, e.g. 1.0 GPa, $k_{\rm F}$$\xi_{//}$ reaches ~50, indicating that thousands of pairs overlap, as in the conventional BCS regime. On the other hand, at low pressures, $k_{\rm F}$$\xi_{//}$ is as small as 3. According to a theoretical suggestion \cite{engelbrecht1997bcs}, the size of the Cooper pair may become smaller than $\xi_{//}$ when approaching the BEC regime. Therefore, only a few pairs overlap at low pressures below 0.4 GPa, indicating more BEC-like Cooper pairing than higher pressures. This pressure dependence is viewed as a Mottness-driven BEC-to-BCS crossover. 

\subsection{How does the doped quantum spin liquid form superconductivity?}
Taken together with the discussions in this review, $\kappa$-Hg$_{2.89}$Br$_8$ is suggested to host a doped QSL that exhibits BEC-like superconductivity at low pressures. The charge transport, which is confined in the layer at high temperatures, gets interlayer phase coherence below 20 K and condensate into a superonducting state below 6 K as discussed in Chapter 5. Thus, the behavior of the phase of their wave function is a key to the superconducting condensate.

In the vicinity of the superconducting transition under a magnetic field, flowing vortices or preformed Cooper pairs are accompanied by phase fluctuations of the wave functions. Such superconducting fluctuations are known to cause the Nernst signal e$_N$ (=$E_y$/$\bigtriangledown_xT$), which is an electric field generation $E_y$ perpendicular to a thermal gradient $\bigtriangledown_xT$ and magnetic field $B_z$, as observed in cuprates, iron chalcogenides, and organic conductors \cite{wang2006nernst,kang2020preformed,nam2007fluctuating,nam2013superconducting}. A Nernst effect is analogous to a Hall effect in terms of transverse voltage generation under a certain flow, and these coefficients are associated with the Mott relation when quasiparticles mainly contribute to both entropy flow and charge flow \cite{behnia2016nernst, oike2022topological}. The magnetic-flux flow bring about an entropy flow but is not necessarily accompanied by charge flow, and can contribute to the Nernst effect. Thus, the Nernst effects can possibly capture superconducting fluctuations in a more sensitive manner than the Hall effects.

%%%%%%%%%%%%%%%%%%%Fig19
\begin{figure}
\includegraphics[width=86mm]{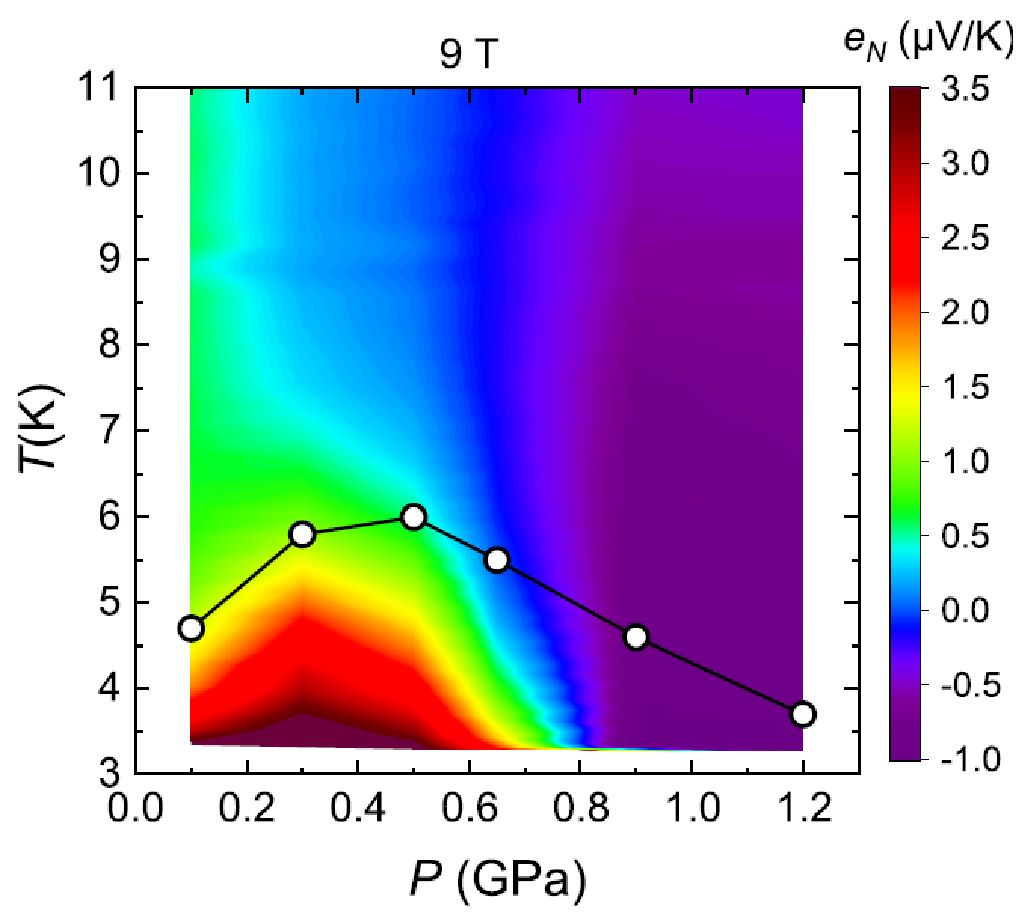}
\caption{\label{fig}
Temperature-pressure ($T$-$P$) profile of the Nernst signal $e_{\rm N}$ at a magnetic field of 9 T. The white circles indicate superconducting transition temperatures at $H=0$. Reprented from Ref. \cite{suzuki2022mott} under the creative commons license.}
\end{figure}
%%%%%%%%%%%%%%%%%%%Fig19

Figs. 18(a)-18(d) display the magnetic-field dependences of e$_N$ at several fixed temperatures under pressures of 0.3, 0.5, 0.65, and 0.9 GPa, together with their contour plots (Figs. 18(e)-18(h)) \cite{suzuki2022mott}. The very sensitive pressure dependence of e$_N$ profile is evident; at 0.3 GPa, the $e_{\rm N}$ value remains large even at the maximum field, 9 T, whereas it rapidly drops above 1 T at 0.9 GPa. Accordingly, the contour line of $e_{\rm N}$ is almost vertical above 3 T at 0.3 GPa, and gradually inclined with increasing pressure. When the $e_{\rm N}$ profiles are compared with the $\rho_{//}$profiles in Fig. 16, the location of of large values of $e_{\rm N}$ in the temperature-pressure plane roughly  coincides with those of $H_{c2\perp}$ determined by $\rho_{//}$/$\rho_{\rm n}$ = 0.9. It is remarkable that, at low pressures, the Nernst signals remain large even at magnetic fields much higher than $H_{c2\perp}$. To highlight the persistence of Nernst signal generations in a nominally normal state, we show, in Fig. 19, a pressure-temperature dependence of $e_{\rm N}$ at 9 T in a contour plot \cite{suzuki2022mott}. The Nernst signals that emerges from well above $T_{\rm c}$ in the low-pressure range are suppressed with increasing pressure, indicating their relevance to Mottness.

The persistence of Nernst signals at high temperatures well above $T_{\rm c}$ is among the properties of BEC-like condensates; preformed pairs are expected to generate the Nernst signal in a more sensitive manner than the resistivity \cite{chen2005bcs}. It has been reported that the $^{13}$C NMR relaxation rate divided by temperature, 1/$T_1T$, takes a maximum value upon cooling at a temperature of 7–9 K, which is twice as high as $T_{\rm c}$ of 4.2 K, at ambient pressure, and this peak formation becomes less prominent with increasing pressure \cite{eto2010non}. The recent torque measurements suggest that the fluctuating superconductivity sets in at approximately 7 K at ambient pressure \cite{imajo2021extraordinary}. These behaviors are also likely to capture the preformed pairs in the BEC-like regime at ambient pressure and its crossover to the BCS regime at high pressures. Upon cooling across $T_{\rm c}$, the preformed pairs turn to a vortex liquid state as manifested by a decrease in resistivity and the large Nernst signals. The field-robust vortex liquid state at low pressures is also consistent with the small values of $k_{\rm F}$$\xi_{//}$ inherent in BEC-like pairing, as shown theoretically \cite{adachi2019stabilization}. Thus, a doped QSL presumably hosts preformed pairs in a nominally normal state and becomes superconducting when their phase coherence grows sufficiently at low temperatures.

\section{Summary}
In this review, we describe what the study of $\kappa$-Hg$_{2.89}$Br$_8$ has opened up in the physics of strongly correlated electron systems. The pressure variation in the resistivity and Hall coefficient suggests that a bandwidth-controlled Mott-like transition, called Mottness transition here, occurs even in the condition that the band filling deviates from a half. Unlike the metal-insulator transitions in half-filled systems, pressure variation of Mottness in the doped system brings about two kinds of metallic states: a non-Fermi liquid and a conventional Fermi liquid, as described in Chapter 4. In the non-Fermi liquid regime, the spin susceptibility of $\kappa$-Hg$_{2.89}$Br$_8$ behave almost identically to the quantum spin liquid (QSL) in the Mott insulator $\kappa$-(BEDT-TTF)$_2$Cu$_2$(CN)$_3$ when scaled by the exchange interactions. The invariant QSL-like behaviors irrespective of whether it is metal or insulator demonstrates a spin-charge separation driven by Mottness. The charge transport is confined in a layer at high tempeartures, but gains interlayer coherence at low temperatures, forming unconventionally degenerative excitations, as described in Chapter 5. Out of the degeneracy, preformed pairs gradually appear on cooling as seeds of superconductivity, as indicated by the enhancement of Nernst effects, and eventually, a superconducting transition occurs. The upper critical field of superconductivity strongly depends on pressure, and the deduced inter-Cooper-pair distance uncovered a drastic pressure variation in superconducting nature from Bose-Einstein Condensation-like to Bardeen-Cooper-Schrieffer-like, as described in Chapter 6. Thus, we traced the energetic hierarchy of $\kappa$-Hg$_{2.89}$Br$_8$ in the order of Mottness, spin liquidity, and superconductivity, which occur in charge, spin, and charge-spin coupled degrees of freedom, respectively. At each energy scale, correlated electrons exhibit unconventional behaviors and the pressure control of correlation reveals how the unconventional states transition into the conventional metal and superconductor.

In this review, we discussed the physics behind the experimental results particularly focusing on transport, thermoelectric and magnetic properties. Although experimental methods available under pressure are limited, the extension of these measurements to much higher magnetic fields or spectroscopic studies are expected to afford deeper insight into the interplay between QSL and charges or unveil novel phenomena yet to be observed in strongly correlated systems so far. On the other hand, we anticipate that the development of synthesis methods that produce pressure-controllable doped Mott insulators other than $\kappa$-Hg$_{2.89}$Br$_8$ will widen the present discussion based on a specific material to a broader perspective. Moreover, a comprehensive understanding of correlated electrons should require integration of concepts across multiple research areas, for example, as we compared the present single-band system with heavy electron systems with multiband. It may be a promissing way in this line to attempt to find physical linkage among the diverse phenomena in various materials. We hope this review will help find cross-disciplinary concepts for correlated electrons.

\section*{ACKNOWLEDGMENTS}
	The authors thank Yuji Suzuki, Kodai Wakamatsu, Jun Ibuka and Takenori Fujii for their significant contributions to the studies related to this review, and Dita Sari Puspita for the picture of a large single crystal. This work was supported by JST PRESTO (Grant No. JPMJPR21Q2), and JSPS KAKENHI (Grants No. 11J09324, No. 22H01164, No. 21K03438, 18H05225, 21K18144, 20K20894, 20KK0060).

\bibliography{Oike_ref}
%\begin{thebibliography}{99}
%\end{thebibliography}

\end{document}